\documentclass[twocolumn,showpacs,preprintnumbers,nofootinbib,amsmath,amssymb]{revtex4}
\usepackage{epsfig}
\usepackage{dcolumn}
\usepackage{bm}

\newcommand\al{\alpha}
\newcommand\be{\beta}
\newcommand\de{\delta}

\newcommand\De{\Delta}

\newcommand\half{{\frac{1}{2}}}
\newcommand\tralf{{\frac{3}{2}}}
\newcommand\trelf{{\frac{3}{4}}}
\newcommand\fotrelf{{\frac{4}{3}}}

\newcommand\ce{{\cal E}}

\newcommand\ep{\epsilon}
\newcommand\vep{\varepsilon}
\newcommand\MD{\mathfrak{D}}
\newcommand\BMD{\bar{\mathfrak{D}}}

\begin{document}
\title{Three-body breakup within the fully discretized Faddeev equations}

\author{O.A. Rubtsova }%
\email{rubtsova-olga@yandex.ru}
\author{V.N. Pomerantsev}
\email{pomeran@nucl-th.sinp.msu.ru}
\author{V.I. Kukulin}
\email{kukulin@nucl-th.sinp.msu.ru}

\affiliation{%
Skobeltsyn Institute of Nuclear Physics, Lomonosov Moscow State University, 
Leninskie Gory 1(2), 119991  Moscow, Russia}

\author{Amand Faessler}
 \email{faessler@uni-tuebingen.de}
\affiliation{%
Institute for Theoretical Physics, University of T\"ubingen, Auf der
Morgenstelle 14, D-72076 T\"ubingen, Germany}

\date{\today}

 \begin{abstract}
A novel approach is developed to find the three-body breakup amplitudes and
cross sections within the modified Faddeev equation framework. The method is based on
the lattice-like discretization of the three-body continuum with a three-body
stationary wave-packet basis in momentum space. The approach makes it possible
to simplify drastically all the three- and few-body breakup calculations due to
discrete wave-packet representations for the few-body continuum and simultaneous
lattice representation for all the scattering operators entering the integral
equation kernels. As a result, the few-body  breakup can be treated as a
particular case of multi-channel scattering in which part of the channels
represents the true few-body continuum states. As an illustration for the
novel approach, an accurate calculations for the three-body breakup process
$n+d\to n+n+p$  with non-local and local $NN$ interactions are calculated.
The results obtained reproduce nicely the benchmark calculation results
using the traditional Faddeev scheme which requires much more tedious and
time-consuming calculations.
\end{abstract}
 \maketitle

\section{Introduction}
The last decades have inaugurated great success in precise
{\em ab-initio} calculations for few-body scattering processes
\cite{Gloeckle_rep,F2,Elster,Fonseca,F1,Sauer,Kievsky}. These calculations
made it possible to describe accurately the results of numerous
recent experiments on elastic $nd$ scattering at energies up
to 350 MeV and also the three-body breakup $n+d\to n+n+p$ at low and
moderate energies $E_{n}\simeq 10\div30$ MeV. However some problems
remain unsettled even at such low energies. These are the so-called
$A_y$-puzzle (as well as other puzzles for various tensor and
vector analyzing powers) in elastic scattering, the problems
with an adequate description of the pairwise $^1S_0$-channel
contribution to three-body breakup at low energies
\cite{Gloeckle_nn} and breakup cross section in some particular
three-particle configurations such as the quasi-free scattering
\cite{Ruan} and the space-star \cite{S_star} configurations. The most
plausible reason for the visible discrepancies with experimental
data in this area is likely not insufficient accuracy of numerical
calculations but rather some deficiency in the input   $2N$- and
$3N$-interactions. At the same time, the progress in the field of
precise few-nucleon calculations, particularly in testing of new
models for $3N$-interactions, is restrained strongly by a high
number of very complicated
few-nucleon calculations, especially above the
three-body threshold. Because of these complications  of traditional
computational schemes for the direct solution of the Faddeev--Yakubovsky
equations, one observes a rise of interest in recent years in alternative approaches
\cite{Carbonell,LIT,Gl_Raw} to calculate  the scattering observables by simpler methods.

Among such alternative approaches one can note a preference for
the so called $L_2$-methods. These methods are based on
expansions of the scattering solution into a  basis of
square-integrable functions
\cite{levin,Kurouglu,mccurthy,horse,papp,Piyadasa,Rawitscher,CDCC_br}.
Such $L_2$-methods proved to be very well suited and quite efficient
for numerous applications. One of the most successful approaches of
this type is the Continuum-Discretized Coupled-Channel (CDCC) method
developed three decades ago for treatment of breakup processes in
direct nuclear reactions
\cite{Piyadasa,Rawitscher,CDCC_br,Thompson}. The CDCC approach in
its traditional form was unable to treat other channels than
elastic scattering and projectile (or  target) breakup. Recently a few groups
generalized the traditional CDCC approach to scattering of three-fragment
projectiles by a stable target \cite{CDCC_br}. However this
generalized approach can  be considered as a hybrid method: $L_2$
discretization of inner motion in the three-body projectile  and the
traditional treatment of a coupled-channel problem

On the other hand, the present authors have developed some
alternative $L_2$-technique \cite{Moro,KPR1,KPR2,K2} which is based
on the idea of complete continuum discretization with a
special stationary wave-packet basis in momentum space (three-body
lattice basis). The basic distinction of such an approach from the
traditional CDCC-scheme for the three-body systems is 
that the wave-packet approach is dealing with a full discretization of
the three-body continuum. In other words, the discretization on
{\em both} Jacobi coordinates is used here rather than the discretization on the
alone coordinate of the projectile inner motion as in the
CDCC approach\footnote{It is interesting to add that similar idea of
global three-body discretization in a momentum space has been
proposed earlier \cite{Kurouglu} within the pseudostate extension of
the coupled-reaction-channel method. The author solved as an
illustration of the approach the simple model problem of 2$\to$2
scattering and also the breakup $2\to3$ process using the Laguerre
polynomial basis. Unfortunately this prospective approach has not
received any further development.}.

Our approach with the global discretization over all valence
coordinates leads immediately to a few important advantages in the
accurate treatment of few-body scattering. Among those
the following are the most important:
\begin{itemize}
\item[(i)] The few-body scattering problem is consistently formulated in a
Hilbert space of three-body {\em normalized} states, similarly to the
bound-state problem.
\item[(ii)] The approach employs the integral equation framework of scattering
theory instead of the differential equation approach (e.g. in the
CDCC) where the boundary conditions in few-body scattering channels
are not easy to formulate, especially in terms of the $L_2$ basis
used. Contrary to this, the integral equation formulation allows to
avoid any explicit account of the boundary conditions.
\item[(iii)] When working within the wave-packet formulation of the scattering
problem one can derive explicit formulas for some scattering
operators (e.g. channel resolvents). Such fully analytical
finite-dimensional approximations, being substituted into integral
equation kernels, lead immediately to their algebraic matrix
analogues. Thus, our final equations are simple matrix linear
equations with regular matrices.
\end{itemize}

In our previous works \cite{KPR1,KPR2} we have demonstrated
how to find the elastic $2\to2$ scattering amplitudes in lattice
representation. In the present paper, we generalize the technique to
the three-body breakup treatment. So we present here the complete
formalism for determination of three-body breakup amplitudes. It is
important to emphasize in this connection that the accurate
treatment of three-body breakup within the $L_2$ type approach is
much less obvious than that of elastic ones and thus requires some
additional delicate theoretical studies. In particular, the matrix
elements in the breakup amplitude are not  truncated over all
spatial coordinates (in contrast to the elastic and rearrangement
amplitudes), so the validity of the  $L_2$ scheme in the treatment of the
breakup processes should be studied carefully. As some
substantiation for such approach one can consider the three-body breakup
calculations within the CDCC-approach where the discretization of
 the continuum in the projectile inner subHamiltonian has been used
 for the description of the breakup amplitudes
\cite{CDCC_br,Thompson}. So, the natural generalization of such
a partial continuum discretization to the case of full three- and
few-body continuum within the Faddeev equation approach is an
important next step. Moreover, this fully discretized approach
studied in the present work  allows to simplify drastically all
calculations and makes it more universal and elegant.

The present work has the following structure. In the section II, a
three-body lattice-like free wave-packet basis is described in
detail  together with a similar basis  for the channel
Hamiltonian. Here we also discuss the properties of these bases. The
complete formalism for elastic scattering and breakup, as applied to the
$nd$-system in the packet representation is presented in Section
III. In the  Section IV, a few useful numerical illustrations and
their comparison with the  standard Faddeev benchmark calculations
are given. Our results are summarized in the Section V. For the sake
of convenience  for the reader we add three appendices. In Appendix
A we describe the detailed scheme for calculation of the three-body
overlap matrix in the three-body lattice basis for recoupling of
different  Jacobi coordinates. In Appendix B we give  the convenient
wave-packet formalism for the solution of three-body scattering
problem with separable pairwise interactions. In the last Appendix C
we discuss some features of our numerical calculations.

\section{Lattice representation for the three-body continuum}

We consider here the  problem of a scattering of three identical
particles 1, 2 and 3 (nucleons) with mass $m$, interacting via pairwise
short-range potentials $v_a$ $(a=1,2,3)$. It is convenient to use
three Jacobi momentum sets $({\bf p}_a,{\bf q}_a)$ corresponding
to  three channel Hamiltonians $H_a$ $(a=1,2,3)$ which  define
 the asymptotic states of the system. For example, the  channel
Hamiltonian $H_1$ has the form of the direct sum of two-body
subHamiltonians
\begin{equation}
\label{ch_ham} H_1\equiv h_1\oplus h_{0}^1,
\end{equation}
where subHamiltonian $h_1$ describes $NN$ subsystem consisting of
particles 2 and 3 with interaction $v_1$ and subHamiltonian
$h^{01}$
 corresponds to the free motion of nucleon 1 relative to center of mass
 of the  subsystem \{23\}.

As we study the identical particle system, we will omit further,
where it is possible, the  Jacobi  index $a$.

\subsection{The two-body free wave-packet states}
We start from the free-motion three-body Hamiltonian defined in the
given Jacobi momentum set ({\bf p,q})
\begin{equation}
H_0=h_0\oplus h_0^1,
\end{equation}
where the subHamiltonian $h_0$ defines the free motion of two nucleons with the
relative momentum $\bf p$  and the subHamiltonian
$h_0^1$ defines the free motion of the third nucleon with the
 momentum $\bf q$ relative to the center of mass of
given $NN$ subsystem.

Now we will construct our three-body $L_2$ basis using
discretization of the continua of the two above
subHamiltonians. In doing this we will employ the complete sets of
continuum  wave functions $|p\rangle$ and
$|q\rangle$ (for every partial wave)
normalized according to the conditions:
\begin{equation}
\label{dek} \langle
p|p'\rangle=\de(p-p'), \quad
\langle q|q'\rangle=\de(q-q').
\end{equation}

When discretizing, we truncate the continuum of
$h_0$ and $h_0^1$ by maximal values $\ep_{\rm max }$  and $\ce_{\rm
max}$ respectively, so that the continuous spectra above these
values can be neglected. Further, the selected energy regions
$[0,\ep_{\rm max}]$ and $[0,\ce_{\rm max}]$ are divided onto
non-overlapping bins $\{[\ep_{i-1},\ep_i]_{i=1}^M\}$ and
$\{[\ce_{j-1},\ce_j]_{j=1}^N\}$. Such energy bins correspond to
momentum bins $[p_{i-1},p_i]$ and $[q_{j-1},q_j]$, so that the
end-points of both sets are interrelated by conventional formulas
$p_i=\sqrt{m\ep_i}$ and $q_j=\sqrt{{\frac43 m \ce_j}}$. To further
simplify the notation, we will denote the intervals in the variable
$p$ (both the energy and momentum ones) as $\MD_i$ and in the
variable $q$ as  $\BMD_j$. We use also the following notations
for the widths of the corresponding momentum intervals:
 \begin{equation}
d_i=p_i-p_{i-1},\quad \bar{d}_j=q_j-q_{j-1}
\end{equation}

Now let's define a set of {\em  free} stationary wave packets (WPs)
as integrals of the plane waves (corresponding to the {\em free}
motion) over the above momentum bins for both
subHamiltonians\footnote{Below we will use the Gothic letters to
denote objects (wave functions and operators) in the WP subspace.}:
\begin{eqnarray}
\label{ip}
|\mathfrak{p}_i\rangle=\frac{1}{\sqrt{A_i}}\int_{\MD_i}dp
f(p)|p\rangle,\quad i=1,\ldots,M,\\
\label{iq}
|\mathfrak{q}_j\rangle=\frac{1}{\sqrt{B_j}}\int_{\BMD_j}dq
w(q)|q\rangle,\quad j=1,\ldots,N.
\end{eqnarray}
 where $f(p)$ and $w(q)$ are some known weight functions and $A_i$
and $B_j$ are normalization factors, directly related to the
weight functions
\begin{equation}
\label{norm} A_i=\int_{\MD_{i}}dp |f(p)|^2,\quad
B_j=\int_{\BMD_j}dq|w(q)|^2,
\end{equation}
so that the WP states are normalized to unity:
\begin{equation}
\label{normWP}
\langle\mathfrak{p}_i|\mathfrak{p}_{i'}\rangle=
\delta_{ii'},\quad
\langle\mathfrak{q}_j|\mathfrak{q}_{j'}\rangle=\delta_{jj'}.
\end{equation}
It is important to stress, these WP states belong to {\em a Hilbert
space} (similarly to the bound state functions) and WP functions are
square-integrable: in configuration space they vanish at infinity in
contrast to the initial plane waves. But in the relevant restricted range of
configuration space the WP states still resemble quite closely the
exact scattering states taken at the bin center energy (or momentum)
\cite{KPR1}.

The sets
of such WP states $|\mathfrak{p}_i\rangle_{i=1}^M$ and
$|\mathfrak{q}_j\rangle_{j=1}^N$ form  an orthonormalized
bases in Hilbert space
which can be used as  normal $L_2$ bases, e.g. also for variational calculations.

 In our previous papers \cite{KPR1,KPR2} we
have discussed the properties of WP's in detail. A distinctive
feature of  WP bases is that the matrices of the subHamiltonians
found in such bases are diagonal:
\begin{equation}
\langle\mathfrak{p}_i|h_0|\mathfrak{p}_{i'}\rangle=\ep_i^{*}\de_{ii'},\quad
\langle\mathfrak{q}_j|h_0^1|\mathfrak{q}_{j'}\rangle=\ce_j^{*}\delta_{jj'},
\end{equation}
where values $\ep_i^{*}$ and  $\ce_j^{*}$ are defined via
corresponding end-points of bins $\MD_i$ and $\BMD_j$ \cite{KPR1}.
The most useful property of WPs is that the matrices of the
resolvents $g_0(\ep)=[\ep+i0-h_0]^{-1}$ and
$g_0^1(\ce)=[\ce+i0-h_0^1]^{-1}$ are diagonal in the corresponding WP
bases and their elements have explicit analytical forms \cite{KPR1}.

Different choices of  weight functions lead to different sets of
WPs. In practical calculations in this work we use the momentum wave packets
with the unit weight functions:
\begin{equation}
f(p)=1,\ A_i=d_i,\ w(q)=1,\ B_j=\bar{d}_j.
\end{equation}

It is easy to see that the overlap of such free momentum WP  with a
plane wave, i.e. the momentum representation of
packet state (\ref{ip}) itself has the form:
 \begin{equation}
 \label{proj_rule}
 \langle p|\mathfrak{p}_i\rangle=\frac{\vartheta(p\in
 \MD_i)}{\sqrt{d_i}},
 \end{equation}
where  we have introduced a function $\vartheta(p\in \MD_i)$ which
is equal to unity if the momentum $p$ belongs to the interval
$[p_{i-1},p_i]$ and vanishes in the other case. So, the wave packet
$|\mathfrak{p}_i\rangle$ takes a form of simple step-like function
in the momentum representation.

The sets of the constructed free WP states  can be used to find two-body
bound-states and also to solve a two-body scattering problem, e.g. for finding the
two-body off-shell $t$-matrix \cite{KPR1}. In the present paper we will use these
two-body $L_2$ bases to construct three-body WPs for solution of the three-body
scattering problem.


\subsection{Three-body lattice basis and the permutation matrix}

Three-body wave packet states  are built as direct products of
two-body ones. However, here one should take into account the spin
and angular parts of the functions. The total three-body WP basis
function can be written as:
\begin{equation}
\label{xij}
|X_{ij}^{\Gamma\al\be}\rangle=|\mathfrak{p}_i^\alpha\rangle\otimes|\mathfrak{q}_j^\beta\rangle
|\alpha,\beta:\Gamma\rangle,
\end{equation}
where $|\alpha\rangle$ is the  spin-angular state of the $NN$ pair,
$|\beta\rangle$  is the  spin-angular state of third nucleon, while
$|\Gamma\rangle$ is the set of the three-body quantum numbers.
The state (\ref{xij}) is a WP analog of the
exact state of the three-body continuum $|p,q\rangle|\alpha,\beta:\Gamma\rangle$ for the
free Hamiltonian $H_0$.

The properties of such three-body WP's are very similar to those of
two-body wave packets \cite{KPR1}. In particular, the matrix
of the three-body  free Hamiltonian $H_0$ and its   resolvent
$G_0(E)=[E+i0-H_0]^{-1}$ are diagonal in the so constructed basis.
In other words, such a WP basis defines  {\em an ``eigen'' wave-packet
subspace} for the free three-body Hamiltonian $H_0$.

Since the basis functions are the products of both step-like
functions in variables $p$ and $q$,  the solution of the three-body
scattering problem in such a basis corresponds to a formulation of the
scattering problem  {\em on a two-dimensional momentum lattice}.
Therefore we will refer to such a basis  as a {\em lattice basis}. Let us denote the two-dimensional bins
(i.e. the lattice cells) as $\MD_{ij}=\MD_i\otimes\BMD_j$. In the
few-body case, the lattice basis functions are constructed as direct
products of the two-body free WPs, so the basis space
corresponds to a multi-dimensional lattice.

In principle, using the above lattice basis, one can solve a general
three-body  scattering problem by projecting  all the scattering operators onto
such a basis. In particular,  the matrix of the three-body free resolvent  $G_0$
can be expressed in the above lattice representation
fully analytically \cite{K2}.

Let us consider the particle
permutation operator  $P$ which enters in the Faddeev equation for three
identical particles and is defined as
\begin{equation}
P=P_{12}P_{23}+P_{13}P_{23}.
\end{equation}

The matrix of the operator $P$ in the lattice basis corresponds to
the overlap between basis functions defined in different Jacobi
sets:
\begin{equation}
\langle
X_{ij}^{\Gamma\al\be}|P|X_{i'j'}^{\Gamma\al'\be'}\rangle=
\langle
X_{ij}^{\Gamma\al\be}(1)|X_{i'j'}^{\Gamma\al'\be'}(2)\rangle,
\end{equation}
where the argument  1 (or 2) in the basis functions means a
corresponding Jacobi set. Such matrix element can be calculated with
the definition of the basis functions in momentum space
 (\ref{proj_rule}):
\begin{eqnarray}
 [{\mathbb P}^0]_{ij,i'j'}^{\al\be,\al'\be'}\equiv\langle
X_{ij}^{\Gamma\al\be}|P|X_{i'j'}^{\Gamma\al'\be'}\rangle=\nonumber\\
\label{perm}\int_{\MD_{ij}}dpdq\int_{\MD'_{i'j'}}dp'dq'
\times\frac{P^{\Gamma}_{\al\be,\al'\be'}(p,q,p',q')}{\sqrt{d_id_{i'}\bar{d}_j\bar{d}_{j'}}},
\end{eqnarray}
where the prime at the lattice cell   $\MD'_{i'j'}$ indicates that
the cell belongs to the other Jacobi set while the
$P^{\Gamma}_{\al\be,\al'\be'}(p,q,p'q')$ is the kernel of particle
permutation operator in a momentum space. This kernel, as is well
known \cite{schmid}, is proportional to the product of the Dirac
delta and Heaviside theta functions. However, due to ``packetting''
(i.e integration over momentum bins in Eq.~(\ref{perm}))
these singularities get averaged over the cells of the momentum
lattice and, as a result, the elements of  the permutation operator
matrix in the WP basis are finite.

Using the above ``packetting'' procedure and the hyperspherical
momentum coordinates, the calculation of the matrix element in
Eq.~(\ref{perm}) can be  done using only a
 one-dimensional numerical integration over the hypermomentum
$K=\sqrt{p^2+\trelf q^2}$. The  technique of this calculation for
the s-wave basis functions is given in Appendix A of the present
paper. The generalization for higher partial waves is
straightforward.


It should be emphasized here that the fixed lattice-like form for the
permutation operator matrix makes it possible to avoid  the complicated  and
time consuming multi-dimensional interpolations of the current
solution when solving the Faddeev equations (in
momentum space) by iterations in conventional approach \cite{Gloeckle_rep,
Elster}. Such
numerous multi-dimensional interpolations at each iteration step
take a big portion of the computational time in practical numerical
procedure. When solving the four-body Yakubovsky equations the
dimension for these interpolations increase and thus the
computational efforts  get even
higher. So, avoiding the very numerous multidimensional
interpolations in each step of the iterations leads to a tremendous
acceleration for all three-body calculations in momentum space.

Thus, the two-dimensional momentum lattice basis  constructed above
can be applied directly to solving the Faddeev equations for the
conventional transition operator   $U$. However, by using the very
convenient form for the spectral representation of the resolvent
operators in the WP basis one can employ  some alternative (but
equivalent) form of the Faddeev equation, which makes it possible
to avoid the time-consuming calculation of the fully
off-shell $t$-matrix at many energies (which requires
to solve very often the Lippmann-Schwinger equations at every energy and for
different spin-orbit channels) and to replace it by calculating the
resolvent of the $NN$ subHamiltonian $h_1$ in the corresponding {\em
scattering} WP-representation. The latter can be made easily by
straightforward one-fold diagonalization of the $h_1$ subHamiltonian
matrix.


\subsection{The  scattering WPs for the subHamiltonian $h_1$}

As has been demonstrated earlier \cite{KPR1,KPR2,K2}
the stationary wave packets can be built not only for free
Hamiltonians but also for perturbed two-body $h_1=h_0+v_1 $ and
three-body channel Hamiltonians $H_1$.

 In case of the $h_1$
subHamiltonian,  its continua $[0,\vep_{\rm max}]$
for every spin-angular configuration $\alpha$ are divided into
separate bins $\{[\vep_{k-1}^\alpha,\vep_k^\alpha]_{k=1}^K\}$ and
one can build the scattering wave packet for every such bin
$\De_k^\alpha\equiv [\vep_{k-1}^\alpha,\vep_k^\alpha]$ in the form
\begin{equation}
\label{zi}
|z_k^\alpha\rangle=\frac1{\sqrt{D_k^\alpha}}\int_{\De_k^\alpha}{\rm
 d}p|\phi_p^\alpha\rangle,
\end{equation}
i.e. as an integral over the exact scattering wave function
$|\phi_p^\alpha\rangle$ on the energy interval $\De_k^\alpha$.
Here we use the unit weight function
 and  $D_k^\alpha$ is the width of interval $\De_k^\alpha$.

It is easy to show that such packet states have the same properties
with respect to  their ``eigen'' Hamiltonian $h_1$ as
 free WPs with respect to the free Hamiltonian $h_0$.
The only difference is that the set of scattering WP's should be accomplished with
the bound state functions of $h_1$ (if they exist). Jointly with the
possible bound state wave functions for the $h_1$ subHamiltonian,
the scattering  WP's form an orthonormalized basis in which both the matrix of the
Hamiltonian $h_1$  and the matrix of its
 resolvent $g_1(\vep)=[\vep+i0-h_1]^{-1}$ are diagonal \cite{KPR1}.

The projection properties for the WP of $h_1$ will be  similar to
those for $h_0$ (\ref{proj_rule}), viz.:
\begin{equation}
\langle \phi_p^\alpha|z_k^\alpha\rangle=\frac{\vartheta(p\in
\De_k^\alpha)}{\sqrt{D_k^\alpha}}.
\end{equation}

\subsection{Pseudostates as approximations for scattering WPs}
At first glance it may appear that the exact scattering WP  basis
  is useless because its construction would require the knowledge
of exact scattering wave functions of the Hamiltonian $h_1$. However, as has
been demonstrated \cite{KPR1}, the properties of the exact scattering WP's
for $h_1$ are
quite similar to those of respective pseudostates obtained by the
diagonalization of the Hamiltonian matrix in some complete $L_2$-basis.
So that, in actual calculations  one can replace the set of WP's $|z_k^\alpha\rangle$
by the set of respective  pseudostates \cite{KPR1}.  Such an
 $L_2$-basis can be used as a very good approximation for the free WP-basis (\ref{ip}).
As a result of such Hamiltonian matrix diagonalization, one gets
a set of pseudostates
\begin{equation}
  \label{exp_z}
  |\bar{z}_k^\alpha\rangle=\sum_{i=1}^M O_{ki}^\alpha|\mathfrak{p}_i^\alpha\rangle,\quad
  k=1,\ldots,M,
  \end{equation}
together with a set of their eigenvalues $\vep_k^{\alpha*}$.

In this paper, we restrict ourselves to $s$-wave  spin-dependent
pair
  interactions only. We assume that there is a  single bound state
$|z_0\rangle$ (deuteron) with binding energy $\vep_0^*$ in the $NN$
spin-triplet channel and there are
  no bound states in the $NN$ spin-singlet channel.

In case of $s$-wave scattering
with $s$-wave
$NN$ interactions, the indices $\alpha$, $\beta$ and $\Gamma$ in
Eq.~(\ref{xij}) include only the
spin quantum numbers. So, below we will use the value of the spin of $NN$-pair
$s=0,1$ instead of index $\alpha$, while the index
$\beta$, which indicates  the
spin value of the third nucleon (i.e. $\half$),  will be omitted
everywhere. The index $\Gamma$ defining the set of quantum numbers for three-body
states is reduced to
the total spin of the three-body system $\Sigma=\half,\tralf$ which in the
$s$-wave case is equal to the total angular momentum of the system and therefore is
conserved.

After the above diagonalization in the spin-triplet channel
($s=1$) one gets a set of  pseudostate functions, the first of which
 $|\bar{z}_1^1\rangle\approx|z_0\rangle$  with  the energy $\vep_0^*$,
 is an approximation for the
deuteron wave function, while the other $M-1$ pseudostates with
energies $\vep_k^{1*}$ are localized in the continuum spectrum and
correspond to scattering WP states for $h_1$. In the spin-singlet
channel there are no $NN$ bound states, so that all functions
$|\bar{z}_k^\alpha\rangle$ in Eq.~(\ref{exp_z}) are approximated by
scattering WP's. It is important to note that as a by-product of our
diagonalization procedure one gets simultaneously the discrete
representation for $NN$ partial phase shifts $\de^s(\vep_k^{s*})$
 for all pseudo-states energies (i.e. in one step!) -- see
the detail in Ref.~\cite{KPRF}.

Since the free WP basis functions (in the momentum space) are
step-like  functions, the  momentum dependence of all functions
expressed via such a basis have a histogram-like form. An example of
the momentum dependence for the bound state (deuteron) function in
such step-like basis in comparison with the exact function for
Yamaguchi triplet $NN$ potential (see  Appendix B) is displayed in
Fig.~{\ref{fig0}}.
\begin{figure}[h]
\centering \epsfig{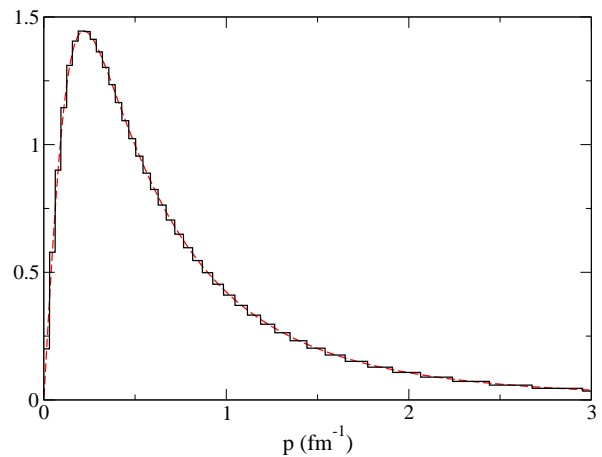}
 \caption{\label{fig0} (Color online) Comparison of the exact deuteron wave
 function obtained in the momentum space for the Yamaguchi potential  
 (dashed curve) with its approximation in the lattice basis (solid line).}
\end{figure}

\begin{figure}[h]
\centering \epsfig{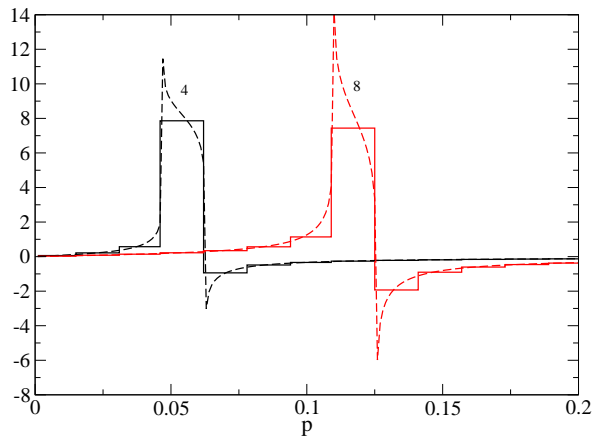}
 \caption{\label{wpac} (Color online) The functions of pseudostates ($k=4,8$) obtained in the lattice
basis (solid lines) in comparison with exact scattering packets
(dashed lines) for the $NN$ spin-triplet Yamaguchi potential. }
\end{figure}

The Fig.~{\ref{wpac}} displays the functions of two pseudostates
(with $k=4,8$) obtained in the lattice basis in comparison with  the
corresponding exact scattering wave packets which can be calculated
exactly for the separable Yamaguchi potential.  It is interesting to
see that although functions of the exact scattering WP's (\ref{zi})
(dashed lines in the figure) have the logarithmic singularities  at
the boundaries of the ``on-shell'' interval (i.e. the one  which the
state energy belongs to)  they are square-integrable as well as the
free-motion WP's.

It is clear from the comparison that the pseudostates composed from
step-like wave packets reproduce very well the structure of the
exact scattering wave packets ``on average''.

Now having at our disposal the two-body bases for subHamiltonians
$h_1$ and $h_0^1$ one can construct the three-body WP basis for the
channel Hamiltonian $H_1$ which defines the asymptotic motion in the
system.

\subsection{Construction of three-body WP basis for the channel Hamiltonian}

The three-body WP states corresponding to the channel Hamiltonian
$H_1$ can be defined similar as the WP-states for the
three-body free Hamiltonian $H_0$, i.e. as direct products of
two-body WP states for $h_0^1$ and $h_1$
subHamiltonians (jointly with the bound-state) multiplied by the spin functions of the system:
  \begin{equation}
\label{si} |Z^{\Sigma s}_{kj}\rangle
\equiv|z_k^s\rangle\otimes|\mathfrak{q}_j\rangle
|s,\half:\Sigma\rangle,
\end{equation}
where $s$ and $\Sigma$ are the $NN$ subsystem and  the total
three-body spins correspondingly\footnote{We consider here the
three-body states with total isospin $T=\half$ only. Since in  case
of $s$-wave $NN$ pairwise interaction the spin- and isospin quantum
numbers are interrelated uniquely by the Pauli principle we can omit
further the isospin parts of the wave functions and corresponding
quantum numbers.}.

 When using the above pseudostate approximation,
these three-body states, similarly to two-body scattering WPs, are
related to the three-body lattice basis states by a simple rotation
transformation (similar to Eq.~(\ref{exp_z})):
\begin{equation}
\label{rotation}
|Z_{kj}^{\Sigma s}\rangle=\sum_{i=1}^MO_{ki}^s|X_{ij}^{\Sigma s}\rangle.
\end{equation}
Hence, starting from the free WP bases for every pair subsystem one
gets a set of basis states both for the three-body free $H_0$ and
the channel $H_1$ Hamiltonians. The basis defined in Eq.~(\ref{si})
defines {\em an ``eigen'' WP-subspace} for the channel Hamiltonian
$H_1$.

This allows us to construct an analytical finite-dimensional
approximation for the channel resolvent $G_1(E)\equiv [E+{\rm
i}0-H_1]^{-1}$.
 Indeed,  the exact three-body channel resolvent  is the convolution of the two-body
 subresolvents $g_1(\vep)$ and $g_0^1(\ce)$:
\begin{equation}
G_1(E)=\frac{1}{2{\pi\rm i}}\int_{-\infty}^{\infty}{\rm d}\vep
g_1(\vep)g_0^1(E-\vep).
\end{equation}
Using further the spectral expansions for the two-body resolvents
and integrating over $\vep$, one gets an explicit
expression for the channel resolvent $G_1$
 as a sum of two terms $G_1(E)=G_1^{\rm BC}(E)+G_1^{\rm
CC}(E)$. Here  the bound-continuum  part $G_1^{\rm BC}(E)$ is the
spectral sum over the three-body  states corresponding to the free
motion of the deuteron relatively to  the third nucleon. So, the
imaginary part of $G_1^{\rm BC}(E)$ is related to a discontinuity on
 the two-body cut of the Riemann  surface of the
three-body energy $E$. The continuum-continuum part   $G_1^{\rm
CC}(E)$ of the channel resolvent includes  the channel three-body
states with the $NN$  pair with interaction in the continuum and ${\rm
Im}G_1^{\rm CC}(E)$ is defined by a discontinuity on the three-body
cut on the energy surface (see the details in Ref.~\cite{KPR1}).

Projecting  the exact channel resolvent onto the three-body channel WP
basis defined in Eq.~(\ref{si}), one can find  analytical formulas
for the matrix elements  of the $G_1$ operator. 
The respective matrix is diagonal in all
wave-packet and spin indices:
\begin{equation}
 \langle Z_{kj}^{\Sigma s}|G_1(E)|
 Z^{\Sigma s'}_{k'j'}\rangle=\de_{kk'}\de_{jj'}\de_{ss'}G^{\Sigma s}_{kj}(E).
 \label{dg1}
\end{equation}

Here the diagonal matrix elements $G_{kj}^{\Sigma s}(E)$ are defined
as integrals over the respective momentum bins and depend in
general on the spectrum partition parameters (i.e. the $p_i$ and
$q_j$ values) and the total energy $E$ only. They do not depend
explicitly on the interaction potential $v_1$.  If the solution of
scattering equations in the finite-dimensional  WP basis converges
with increasing the basis dimension,  the final result turns out to
be {\em independent} on the particular spectral partition
parameters. We have found \cite{K2} the explicit formulas for the
resolvent matrix elements (\ref{dg1}) when one uses  the energy WP's\footnote{%
The matrix elements of the three-body channel resolvent take a simple
analytical form in the WP basis constructed from the continuum wave
functions normalized to the delta-function on the energy (the energy
WPs). For finding the resolvent matrix elements with WP's with
various weight functions one uses  renormalization factors for a
transition from the given wave packet states to the energy ones.},
i.e. WP's with the weight functions $f(p)=\sqrt{p}$, $w(q)=\sqrt{q}$.

 The representation (\ref{dg1}) for the channel resolvent  is the basic expression for
our wave-packet
approach, since it gives explicit analytical formulas for the
three-body resolvent and thus
 it allows to simplify drastically the solution of
general three-body scattering problem. This expression can be used
directly to solve the finite-dimensional analog of the Faddeev equations
for the three
 components of the total scattering wave function~\cite{KPR2}.
Alternatively the very convenient representation (\ref{dg1}) can
also be
 used to solve some particular three-body scattering problems
using the three-body Lippmann--Schwinger equations
\cite{KPR1}.

\section{Solution of $nd$ scattering problem}

Now let us proceed with solving the $nd$ elastic and breakup
scattering problems.

\subsection{The elastic and breakup $nd$ scattering amplitudes} 

The elastic scattering observables can be found from
the Faddeev equation (FE) for the transition operator $\bar U$,
 e.g. in the form \cite{Gloeckle_rep}:
\begin{equation}
\bar{U}=PG_0^{-1}+PtG_0\bar{U},
\end{equation}
where $t$ is two-body $t$-matrix in three-body space and $P$ is the particle permutation operator.
The  equivalent form of FE for the transition operator $U$ has the form:
\begin{equation}
\label{pvg} U=Pv_1+Pv_1G_1U,
\end{equation}
where $G_1$ is the resolvent of the channel Hamiltonian  $H_1$. Since
$tG_0\equiv v_1G_1$ the operators $\bar U$ and $U$ coincide on-shell and
half-shell.

Since the determination of the off-shell channel resolvent in
three-body space is a rather time consuming solution of the FE in the form
(\ref{pvg}), it is very seldom employed for practical solutions.
Actually a similar form of the equations is associated with
 formalisms of the configuration space Faddeev equations,  where
 quite different numerical approaches have been employed \cite{Bound_Gl,Vlach}
than for  the momentum space FE. However, since in the lattice
approach one has explicit analytical  formulas for
the three-body channel resolvent $G_1$ the form (\ref{pvg}) of FE turns out to be
very appropriate for the numerical solution in a WP basis.

The elastic $nd$ scattering amplitude (for a given value of total spin $\Sigma$)
 can be defined as matrix element of the
solution of the Eq.~(\ref{pvg}) taken in the initial state
$|z_0,q_0,\Sigma\rangle$:
\begin{equation}
\label{ael} A^{\Sigma}_{\rm el}(q_0)=\frac23 \frac{m}{q_0}\langle
z_0,q_0,\Sigma|U|z_0,q_0,\Sigma\rangle.
\end{equation}

The breakup amplitude for one Faddeev component of the three-body
wave function (the so-called single-component amplitude) can be found
from the elastic
transition operator $U$ after applying the operator $tG_0$ from
the left:
\begin{equation}
\label{abr} T^{\Sigma s}(p,q)=\frac{ \langle p,q,\Sigma
s|tG_0U|z_0,q_0,\Sigma\rangle}{pqq_0}.
\end{equation}
To obtain the differential breakup cross sections, the total breakup
amplitudes can be found by the contributions of all
three single-component amplitudes.


Thus, we change the conventional treatment of the breakup process \cite{Gloeckle_rep}
and consider the three-body asymptotic states as
scattering states for the channel Hamiltonian $H_1$ rather than as the
states of the three-body continuum for the free Hamiltonian $H_0$. It is a quite
natural when treating the elastic scattering amplitudes because the
initial state wave functions correspond to the channel
Hamiltonian $H_1$. In full analogy with this, one can treat the
deuteron breakup as its excitation into a continuum $NN$-state in the
two-body subsystem governed by the $h_1$  subHamiltonian\footnote{%
It is of interest to remark that while just such a scheme has been used in the
CDCC treatment of breakup processes~\cite{Thompson}, in the Faddeev
approach the final states used for the breakup treatment are the
free three-body states.}.
As was already indicated above such a treatment of breakup processes is close to
the configuration space approach.

In fact, when solving the three-body Faddeev equations in the configuration
space \cite{Bound_Gl,Vlach,GPF} one finds the breakup amplitude
${\cal A}(\theta)$ which determines the asymptotic behavior of the
three-body wave function in hyperspherical coordinates
$\rho=\sqrt{x^2+\frac43
 y^2}$ and $\vartheta=\arctan(\frac{2y}{\sqrt3x})$ as follows
 \begin{equation}
 \psi({\bf x},{\bf y})\xrightarrow[\rho\to \infty]{}\frac{{\cal
 A}(\theta)}{(K\rho)^{5/2}},\quad K=\sqrt{p^2+\trelf q^2},
 \end{equation}
 where $x$ and $y$ are two Jacobi coordinates and $K$ is the hypermomentum.

  This breakup amplitude is defined for every spin-angular configuration and
  interrelated to the partial single-component breakup amplitudes
 (\ref{abr}) by the following formula
\begin{equation}
\label{acal_t} {\cal A}^{\Sigma s}(\theta)=\frac{4\pi
m}{3\sqrt{3}}q_0 K^4 e^{i\pi/4}T^{\Sigma s}(p,q),\
\theta=\arctan(\frac{\sqrt3q}{2p})
\end{equation}
where  $\theta$ is the hyperangle in momentum space.

Now if one transforms the formulas for the breakup amplitudes $\cal
A$ from \cite{Bound_Gl} to the integral form one receives the
following definition for the breakup amplitudes in the momentum
hyperspherical representation
\begin{equation}
\label{acal_t1} {\cal A}^{\Sigma s}(\theta)=\frac{4\pi
m}{3\sqrt{3}}\frac{ K^4}{pq}e^{i\pi/4} \langle
z_0,q_0,\Sigma|U|\phi_p^{s(+)},q,\Sigma\rangle,
\end{equation}
where $|\phi_p^{s(+)}\rangle$ is  scattering function for the Hamiltonian $h_1$
corresponding to the outgoing boundary condition. These
functions are distinguished from the real-valued functions
$|\phi_p^\al\rangle$ used in our approach by only a phase factor:
\begin{equation}
|\phi_p^{s(+)}\rangle=e^{i\delta_s(p)}|\phi_p^s\rangle,
\end{equation}
where $\delta_s(p)$ is the s-wave phase shift of the $NN$
scattering in the channel with spin $s$.

Using formulas
  (\ref{acal_t1}) and (\ref{acal_t}) one can derive an alternative
  to formula (\ref{abr})
  for the {\em single-component} breakup amplitude via
  the scattering functions of the channel Hamiltonian $H_1$
\begin{equation}
\label{t_sig} T^{\Sigma s}(p,q)={e^{i\de_s(p)}} \frac{\langle
z_0,q_0,\Sigma|U|\phi_p^{s},q,\Sigma\rangle}{pqq_0}.
\end{equation}
Summarizing this derivation one can conclude that the breakup
amplitudes can be defined quite similar to a matrix element for the elastic
scattering transition operator $U$ with replacement of the the
$NN$ bound-state wavefunction with the exact scattering
functions for the $NN$ subHamiltonian.

Having now the required representations for both the elastic and
breakup amplitudes, we will proceed in solving the Faddeev equation in
``eigen'' WP subspace of the channel Hamiltonian $H_1$.

\subsection{Solution of the Faddeev equation in the three-body WP basis}
In our wave-packet approach, all the operators in Eq.~(\ref{pvg}) are
projected onto a three-body wave-packet basis corresponding to the channel
Hamiltonian $H_1$. In other word, every operator, e.g. $U$, is
replaced with its finite-dimensional WP representation:
\begin{equation}
\label{umatr} \mathfrak{U}^\Sigma=\sum_{s,kj}\sum_{s',k'j'}
|Z_{kj}^{\Sigma s}\rangle \langle Z_{kj}^{\Sigma s}|U|Z^{\Sigma
s'}_{k'j'}\rangle \langle Z^{\Sigma s'}_{k'j'}|.
\end{equation}

Finally one gets the matrix analog for the Eq.~(\ref{pvg}) (for
the given value of $\Sigma$)
\begin{equation}
\label{m_pvg} {\mathbb U}={\mathbb P}{\mathbb V}_1+{\mathbb
P}{\mathbb V}_1 {\mathbb G}_1 {\mathbb U}.
\end{equation}
Here ${\mathbb V}_1$ and ${\mathbb G}_1$ are the matrices of the
pair interaction and the channel resolvent respectively, the matrix
elements of which can be found in an explicit form.

The matrix ${\mathbb V}_1$ of the potential  $v_1$  is diagonal in the indices
$j,j'$ of the wave-packet basis (\ref{iq}) for the free
subHamiltonian $h^0$ and  has the block form:
\begin{equation}
\label{bv1} [{\mathbb V}_1]^{ss'}_{kj,k'j'}=\de_{jj'}\de_{ss'}
\langle z_k^s|v^s_1|z_{k'}^{s}\rangle.
\end{equation}
These matrix elements do not depend on in index $j$ and can be written with
the usage of the rotation matrix $\mathbb O$ defined in Eq.~(\ref{exp_z}) as:
\begin{equation}
\label{vks}
 \langle z_k^s|v^s_1|z_{k'}^{s}\rangle
 =\sum_{i,i'}O_{ki}^sO_{k'i'}^{s} \langle
\mathfrak{p}_i^s|v^s_1|\mathfrak{p}_{i'}^{s}\rangle\nonumber.
\end{equation}
In the last expression,  the potential matrix elements in the free WP basis are used
which have  the form
\begin{equation}
\langle\mathfrak{p}_i^s|v^s_1|\mathfrak{p}_{i'}^{s}\rangle=
\frac1{\sqrt{d_id_{i'}}}\int_{\MD_i}dp\int_{\MD_{i'}} dp'\
{v_1^{s}(p,p')}
\end{equation}
where $v_1^{s}(p,p')$ is  the momentum representation for
the interaction potential. It implies that the
matrix elements  (\ref{vks}) can be found analytically for a wide variety
of the potential forms.

An important ingredient of our lattice approach presented here is
the representation of the permutation operator $P$ as an overlap
matrix $\mathbb P$ between the  channel WP basis functions for
different sets of the Jacobi coordinates. Using further the
approximation (\ref{exp_z}) for the scattering wave packets
$|z_{k}^s\rangle$, these matrix elements can be expressed through
the overlap matrix ${\mathbb P}^0$  for the free lattice basis
function of Eq.~(\ref{perm}) with the help of the rotation matrices
$\mathbb O$
 \begin{equation}
 \label{perm_z}
\langle Z_{kj}^{\Sigma s}|P|Z_{k'j'}^{\Sigma s'}\rangle \approx
\sum_{ii'}O_{ki}^sO_{k'i'}^{s'*}\langle X_{ij}^{\Sigma
s}|P|X_{i'j'}^{\Sigma s'}\rangle.
\end{equation}

Now let's replace the exact operator $U$ in the formula for the
elastic $nd$ scattering amplitude (\ref{ael}) with its lattice counterpart
$\mathfrak{U}^\Sigma$, and employ  further  the projection rule for the
free WP states (\ref{proj_rule}). Then one gets that the on-shell
elastic amplitude in the wave-packet representation can be
calculated as a a diagonal (on-shell) matrix element of the $\mathbb
U$-matrix:
\begin{equation}
\label{el}  A_{\rm el}^{\Sigma}(E)\approx
\frac{2m}{3q_{0}}\frac{\langle Z^{\Sigma
1}_{0j_0}|\mathfrak{U}^{\Sigma}|Z^{\Sigma
1}_{0j_0}\rangle}{\bar{d}_{j_0}},\quad \Sigma=\half,\tralf,
\end{equation}
where $|Z^{\Sigma1}_{0j_0}\rangle$ is the WP basis state
corresponding to the initial state: index 0 denotes
the bound state of the $NN$ pair (deuteron) and index $j_0$  denotes the
``on-shell'' $q$-bin $\BMD_{j_0}$ with the on-shell momentum
$q_0=\sqrt{\frac{4}{3}m(E-\vep_0^*)}$: $q_0\in \BMD_{j_0}$.

It has been shown above that the breakup amplitude is proportional
to a half-shell matrix element  $\langle z_0,q_0,\Sigma
|U|\phi_p^s,q,\Sigma \rangle$.  Substituting, similarly to the
calculation of the elastic scattering amplitude, the
finite-dimensional operator $\mathfrak{U}^\Sigma$ into the
expression for the breakup amplitude and utilizing the projection
rules for the free-motion and scattering wave-packets, one gets:
\begin{eqnarray}
T^{\Sigma s}(p,q)\approx e^{i\delta(p^*_k)}\frac{{\mathbb T}^{\Sigma
s}_{0j_0,kj}}{p_k^*q_j^*q_0},\nonumber\\
{\mathbb T}^{\Sigma
s}_{0j_0,kj}\equiv\frac{\langle
Z^{\Sigma1}_{0j_0}|\mathfrak{U}^{\Sigma}|Z^{\Sigma
s}_{kj}\rangle}{\sqrt{\bar{d}_{j_0}D_k^s\bar{d}_j}}
,\quad
\begin{array}{c}
q_0\in \BMD_{j_0},\\
q\in \BMD_j,\\
p\in \Delta_k^s,\\
\end{array}
\end{eqnarray}
where $p_k^*=\sqrt{2m\vep_k^{*s}}$ and $q^*_j=\half[q_{j-1}+q_j]$
are momenta corresponding to $\Delta_k^s$ and $\BMD_j$ bins respectively and
the $D_k^s$ is the momentum width of the $\Delta_k^s$ bin.

Thus, we just have found that the elastic and breakup amplitudes can be
calculated directly using the diagonal (``on-shell'') and non-diagonal
(``half-shell'') matrix elements of {\em the same operator}
$\mathfrak{U}^\Sigma$.

However, some problem still arises here: how to define correctly
which of the basis states $Z_{kj}^{\Sigma s}$ correspond to the
on-shell states of the $H_1$.  Due to the  discretization of the
spectrum every WP basis state corresponds to the energy
$E_{kj}^s=\vep_k^{s*}+\ce_j^*$ and thus one does not get the exact
coincidence of these energies for different ``on-shell'' three-body
WP states with the energy  $E$ of the initial state. In other words,
the energy conservation for three-body WP states is fulfilled only
approximately within the corresponding bin widths. To avoid
this difficulty we apply some energy averaging procedure to the
transition matrix elements which is quite natural for the
lattice-like representation.

\subsection{ The energy averaging procedure for the breakup amplitudes.}

Let us to rewrite expression (\ref{t_sig}) for the
breakup amplitude via wave functions of pair subsystems normalized
to $\delta$-functions in energy. Then, when calculating the
breakup amplitudes one will need the transition operator
matrix elements of the form (we omit for the sake of brevity all the spin labels):
\begin{equation}
u(E,\vep)=\langle
z_0,\psi_0(E-\vep_0^*)|{U}|\phi(\vep)\psi_0(E-\vep)\rangle,
\end{equation}
where $|\phi(\vep)\rangle$ is the $NN$ scattering state 
with energy
$\vep$,
$|\psi_0(E-\vep)\rangle$ is the wave function of the 
subHamiltonian $h_0^1$ describing the free motion of the third nucleon
relative to the $NN$ subsystem, and  $E$ is the total energy in the
c.m. system. In the framework of the fully-discretized representation it is quite
natural to make an energy averaging for the transition matrix
elements  $u(E,\vep)$ over the excitation energy  $\vep$, which
leads to the following integrals:
\begin{equation}
\label{fi} u_n(E)\equiv\frac1{\De_n}\int_{\vep_{n-1}}^{\vep_n}d\vep
\langle z_0,\psi_0(E-\vep_0^*)|U|\phi(\vep)\psi_0(E-\vep)\rangle,
\end{equation}
here $\{[\vep_{n-1},\vep_n]\}$ is some set of intervals, in general independent of the
initial partition of the excitation energy $\vep$.
$\De_n=\vep_n-\vep_{n-1}$ are the corresponding widths.

Further, by  replacing the exact  $U$ operator with its wave packet
counterpart (\ref{umatr}) and using the projection rules for the
scattering and free WP's, one can define a new approximation for
the breakup amplitudes:
\begin{eqnarray}
\label{fpe} {\mathbb T}^{\Sigma s}_n(E)= \sum_{kj}{\mathbb
T}^{\Sigma s}_{0j_0,kj}
\frac1{\De_n}\big[\min(\vep_n,\vep_k,E-\ce_{j-1})-\\
-\max(\vep_{n-1},\vep_{k-1},E-\ce_j)\big].\nonumber
\end{eqnarray}
Here the sum runs over all possible indices $k$ and $j$ for which
the difference in the square brackets is positive.

Then, for the single-component breakup amplitudes, one obtains the
following approximate expression with the elements  ${\mathbb
T}_n$:
 \begin{equation}
T^{\Sigma s}(p_n^*,q_n^*)\approx e^{i\delta(p^*_n)} \frac{{\mathbb
T}_n^{\Sigma s}}{p_n^*q_n^*q_0},\quad
  \begin{array}{l}
   p_n^*=\sqrt{{m\vep_n^*}},\\
q_n^*=\sqrt{\fotrelf m(E-\vep_n^*)},\\
\vep_n^*=\half[\vep_{n-1}+\vep_n]. \end{array}
\end{equation}
Now, using the energy averaged WP amplitudes one can get easily
the following formula for the breakup amplitude in the
hyperspherical representation:
\begin{equation}
\label{ca_mt} {\cal A}_n^{\Sigma s}(\theta)=\frac{4\pi
m}{3\sqrt{3}}\frac{K^4}{p_n^*q_n^*}e^{i\delta(p^*_n)}{\mathbb
T}^{\Sigma s}_n,\quad \cos\theta=\sqrt{\frac{\vep_n^*}{E}}.
\end{equation}

\subsection{Breakup differential cross section}

After the determination of the single-component breakup
amplitudes, the total breakup amplitude   is derived using
contributions of all three Faddeev components and can be written
with the help of the particle permutation operator $P$ in the
following way:
\begin{equation}
\label{abr_tot} A_{\rm br}^{\rm tot}({\bf p},{\bf q};{\bf
q}_0)=\langle {\bf p},{\bf q}|(1+P)tG_0U|z_0,{\bf q}_0\rangle.
\end{equation}

The differential breakup cross section is related to this total
breakup amplitudes as follows \cite{Gloeckle_rep}:
\begin{equation}
\label{cross} \frac{d^5\sigma}{d{\bf \hat{k}}_1d{\bf \hat{k}}_2dS}=(2\pi)^4
\frac{2m}{3q_0}\bar{k}_S |A_{\rm br}^{\rm tot}({\bf p},{\bf q};{\bf
q}_0)|^2 .
\end{equation}
Here ${\bf k}_1$ and ${\bf k}_2$ are momenta of particles to be registered,
$S$ is the arclength of the kinematical curve and $\bar{k}_S$ is the phase space
volume defined as:
\begin{equation}
\bar{k}_S=\frac{m^2k_1k_2}{\sqrt{\left(\frac{2k_2-{\bf
\hat{k}}_2({\bf k}_{\rm lab}-{\bf k}_1)}{k_2} \right)^2+
\left(\frac{2k_1-{\bf \hat{k}}_1({\bf k}_{\rm lab}-{\bf k}_2)}{k_1}
\right)^2}},
\end{equation}
where ${\bf k}_{\rm lab}=\tralf{\bf q}_0$ is the center of mass
momentum in the laboratory system.

In the case of $s$-wave $NN$ interactions, the total breakup
amplitude is defined as sum of spin-quartet ($\Sigma=\tralf$) and spin-doublet
($\Sigma=\half$) single-component terms defined for all different Jacobi sets.
E. g. the $S$-wave
amplitude of the two-neutron emission can be represented as a sum of
three terms
\begin{equation}
\label{abr_p} |A_{\rm br}^{\rm tot}|^2=(2|M^{\tralf
1}|^2+|M^{\half 0}|^2+|M^{\half 1}|^2)/3,
\end{equation}
where $M^{\Sigma s}$ are the total amplitudes for the quartet and
doublet channels. In accordance with Ref.~\cite{Cahill}, they are
expressed through single-component breakup amplitudes defined for
different sets of Jacobi momenta  $(p_a,q_a)$ as follows (we assume
here that neutrons are particles 1 and 2 and the proton is the
particle 3):
 \begin{equation}
 \begin{array}{ll}
 M^{\tralf 1}=&T^{\tralf 1}(p_1,q_1)-T^{\tralf 1}(p_2,q_2),\\
 M^{\half 0}=&\frac{2}{\sqrt3}\Big\{\frac14[T^{\half 0}(p_1,q_1)+
 T^{\half 0}(p_2,q_2)]-\\
 &\trelf[T^{\half 1}(p_1,q_1)+T^{\half 1}(p_2,q_2)]+T^{\half 0}(p_3,q_3)\Big\},\\
  M^{\half 1}=&\frac12\Big\{T^{\half 0}(p_1,q_1)-T^{\half
  0}(p_2,q_2)+\\
 &T^{\half 1}(p_1,q_1)-T^{\half 1}(p_2,q_2)\Big\}.\\
 \end{array}
 \label{mdif}
 \end{equation}
Finally the differential cross section of the $n-d$
breakup is expressed through the partial total amplitude
(\ref{abr_p})  as:
\begin{equation}
\label{cross1} \frac{d^5\sigma}{d\hat{k}_1d\hat{k}_2dS}=\frac\pi4
\frac{2m}{3q_0}\bar{k}_S |A_{\rm br}^{\rm tot}|^2.
\end{equation}

To determine the differential cross section for the
two-neutron emission one has to calculate the elements ${\mathbb
T}_n$ (\ref{fpe}) for every spin component $(\Sigma s)$ of the total
breakup amplitude and then substitute them into the explicit
formulas (\ref{mdif}) and (\ref{abr_p}).

Thus, it has been demonstrated above that in our WP approach one can find quite
naturally all breakup amplitudes together with the elastic  scattering amplitude.
This gives a very nice {\em universal} and unified calculation scheme.
Some details of the numerical procedure
for solving the matrix equation~(\ref{m_pvg}) are discussed in Appendix C.

\section{Illustrative examples}

Because we have developed here a novel approach for the treatment of the
three-body breakup processes we would need  precise and reliable tests to
check our new procedure. In all the tests below we employ
as a convenient universal WP basis the free momentum  wave packets for Jacobi
momenta  $q$ and $p$ constructed using the generalized
Tchebyshev grid:
\begin{equation}
\label{q_mesh} q_i=q_{\rm m}\left[\tan\left(
\frac{i}{2N+1}\pi\right)\right]^t, i=1,\ldots,N,
\end{equation}
where $q_{\rm m}$ is the common scale parameter and the $t$-parameter
determines the ``sparseness degree'' of the bin set. A similar
grid with the size $M$ and common scale   $p_{\rm m}$ is introduced
for discretization of the momenta $p$.

\subsection{$nd$ breakup amplitudes for a separable $NN$ potential}
As a first extremely convenient test we have chosen the three-body model with a separable
$NN$-potential. This model, which can be treated numerically very accurately
in various kinematical breakup situations seems to be very
appropriate for such a test (see below).

We consider here the $nd$ breakup with pairwise
separable $NN$-interactions in the form:
\begin{equation}
\label{sep_pot}
v^s=\lambda_{s}|\varphi_{s}\rangle \langle \varphi_{s}|,\quad s=0,1
\end{equation}
As it is well known, the Faddeev equations for such potentials can be
reduced to one-dimensional integral equations in momentum space.
Such equations, as demonstrated in our previous work
\cite{KPR1}, can be solved quite accurately using the
two-body free WP basis. For convenience of the reader we describe all
the details of this procedure in  Appendix B.
Here we will refer to the results found in such approach as to the
``exact ones''. We will compare the latter with the results of the
solution of the general three-body WP scheme with the
Eq.~(\ref{m_pvg}) for the separable potential.

\begin{figure}[h] \centering
{\epsfig{file=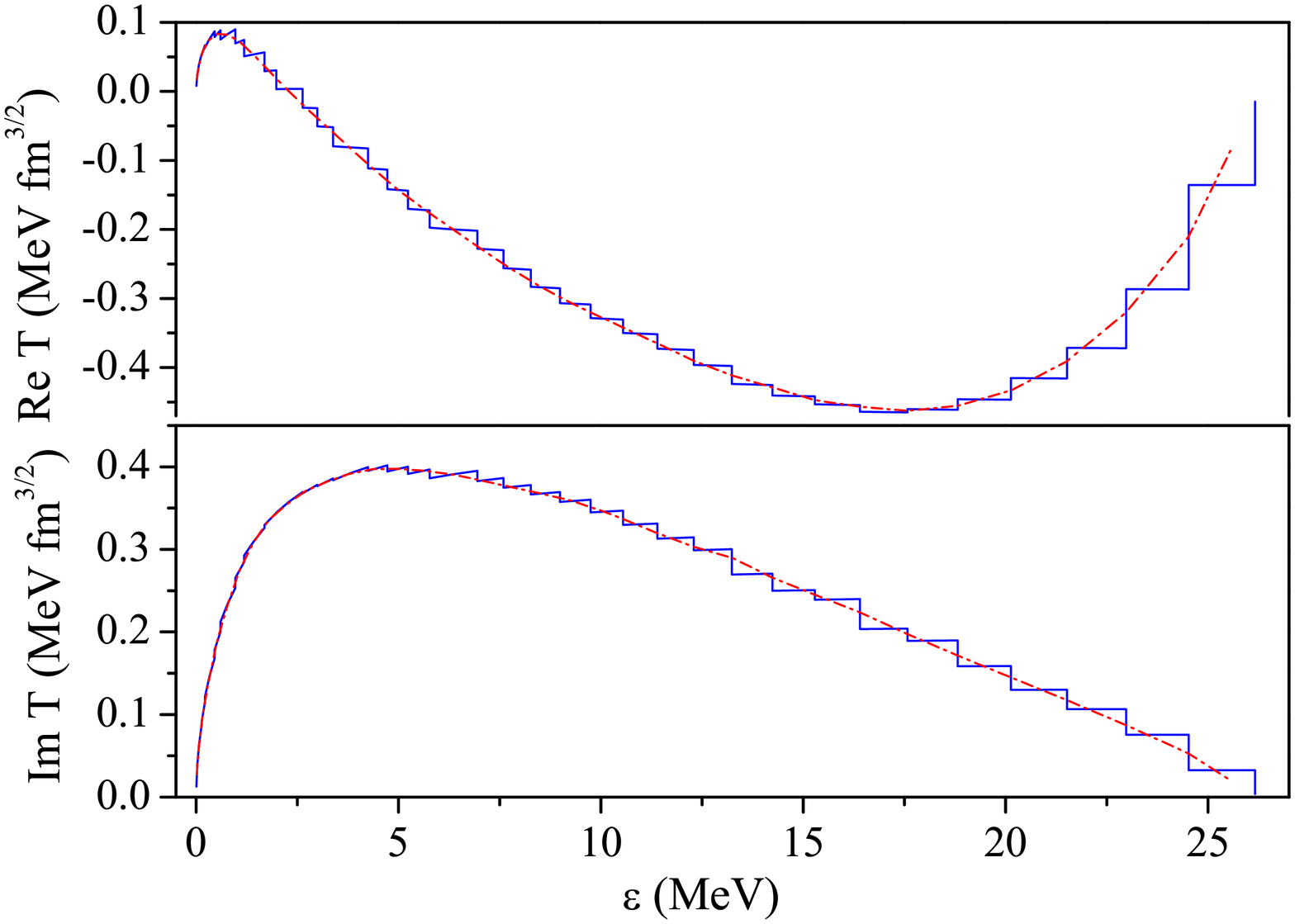,width=\columnwidth}}
 \caption{(Color online) Comparison of the energy unaveraged (solid curve) and
averaged  (dashed-dotted curve) breakup amplitudes ${\mathbb T}^{\tralf 1}$
for the separable   $NN$ potential (\ref{sep_pot}) calculated with
the general WP technique  at the basis size
 � $M=N=100$ . \label{fig02}}
 \end{figure}

The Fig.~\ref{fig02} shows the unaveraged (i.e. having a histogram
form) and energy averaged breakup amplitudes for the Yamaguchi
potential obtained from the general matrix FE (\ref{m_pvg}) in the
lattice basis.  The energy averaged  amplitudes   have been
calculated using the averaging procedure described above. It is
clearly seen from Fig.~\ref{fig02} the method developed makes it
possible to find rather smooth energy dependence for the breakup
amplitudes. Using the averaged amplitudes, we have found the
single-component breakup amplitudes ${\mathbb{T}}^{\Sigma s}$  as
functions of hyperangle $\theta$ for both quartet and doublet
three-body spin channels.

\begin{figure}[h] \centering
{\epsfig{file=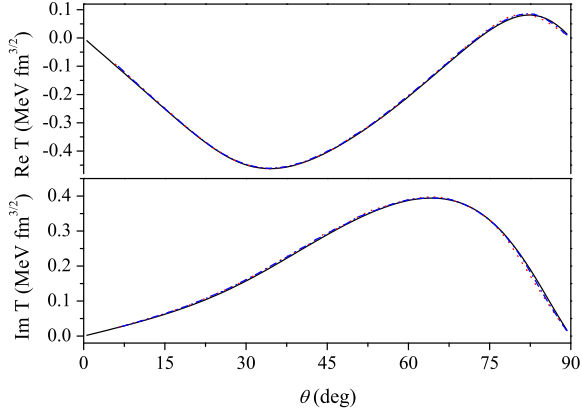,width=\columnwidth}}
 \caption{(Color online) The  breakup amplitude $\mathbb{T}^{\tralf 1}$ in the quartet channel calculated for
the separable  $NN$ potential via the general WP technique for basis
dimensions $M=N=100$ (dotted curves) and $M=N=200$ (dash-dotted
curves)  in comparison with the 'exact' values   (full curves).
with the resolution of this figure the three curves
can practically not be distinguished.
\label{fig1}}
 \end{figure}

\begin{figure}[h] \centering
{\epsfig{file=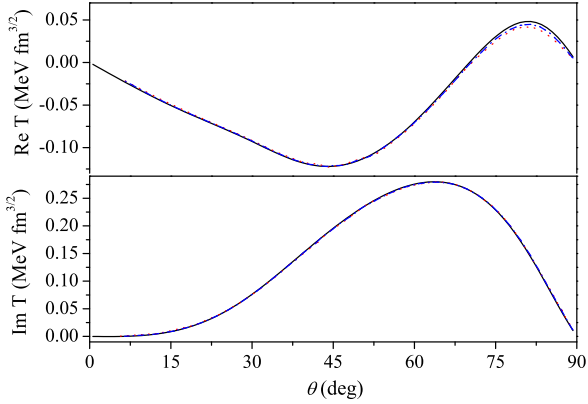,width=\columnwidth}}
 \caption{(Color online) The breakup amplitude $\mathbb{T}^{\half 1}$  in the spin-doublet
 channel for a separable
 $NN$ model. The notations are the same as in the Fig.~\ref{fig1}. \label{fig2}}
 \end{figure}

\begin{figure}[h] \centering
{\epsfig{file=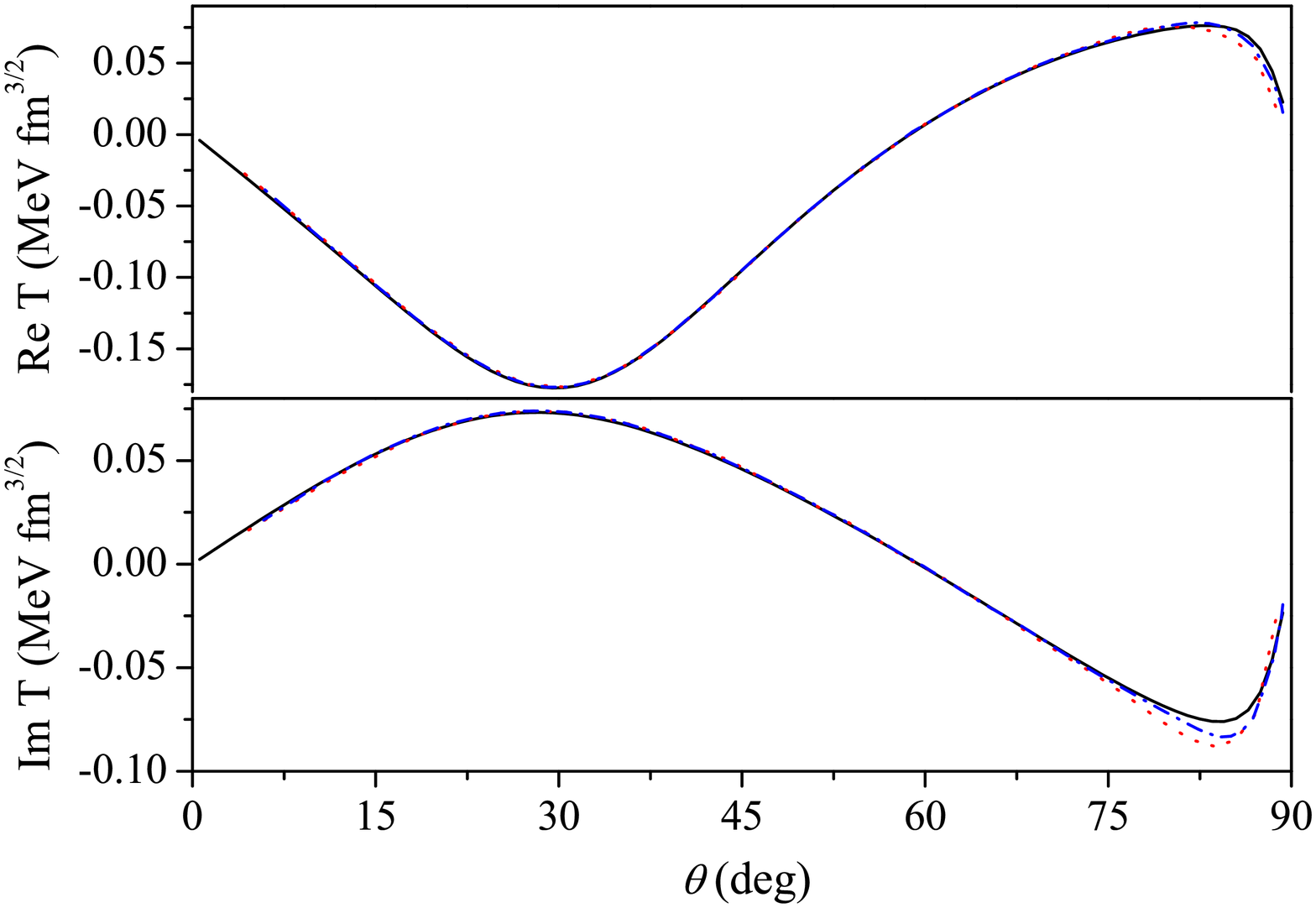,width=\columnwidth}}
 \caption{(Color online) The  breakup amplitude $\mathbb{T}^{\half 0}$  in the spin-doublet
channel for the
separable $NN$-model. The notations are the same as in the Fig.~\ref{fig1}. \label{fig3}}
 \end{figure}

Now, we can compare the approximated WP breakup amplitudes
(\ref{fpe}) found within our general formalism for a separable model
with  the exact amplitudes derived from directly solving the
one-dimensional Faddeev equations. In the
Figs.~\ref{fig1}-\ref{fig3} such a comparison between approximated
and ``exact'' results is presented. The two-body free WP bases with
size $M=200$ and $N=200$  have been used in the calculation of the
approximated amplitudes.

We observe in the Figs.~\ref{fig1}-\ref{fig3} quite a very
good agreement between the lattice-approximations and the ``exact''
amplitudes. The corresponding curves are almost indistinguishable in the
figures. The only differences are seen at
 the hyperangle region $\theta\sim$ 90$^\circ$ for the spin-doublet $T^{\half
0}$ amplitude. It is clear also that the WP amplitudes calculated in a
finite-dimensional $L_2$ basis
 converge to the exact ones with increasing basis size.

\subsection{n-d breakup amplitudes for a local  NN potential}

Having tested our novel approach using the simple separable model for the
$NN$ force one can move to a more realistic case of a local
$NN$-interaction. For this, we have chosen the so-called
 MT I-III $NN$ central potential which was frequently used in the past for the test
of few-body calculations. So, we can compare our results for this model
with very accurate benchmark calculations  \cite{Friar}. For the present WP
calculations we use again the three-body lattice basis constructed
on a Tchebyshev two-dimensional grid.

\begin{figure}[h] \centering
{\epsfig{file=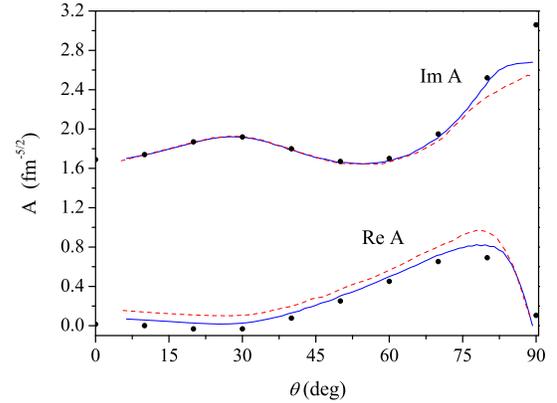,width=\columnwidth}}
 \caption{(Color online) The breakup amplitudes ${\cal A}^{\tralf 1}$ in the spin-quartet channel calculated
using the WP technique for the $NN$ force MT I-III for basis
size $M=N=100$ (dashed curves) and $M=N=200$ (solid curves) in
comparison with the results of the benchmark calculations (solid
circles) \cite{Friar}. \label{fig4}}
 \end{figure}

\begin{figure}[h] \centering
{\epsfig{file=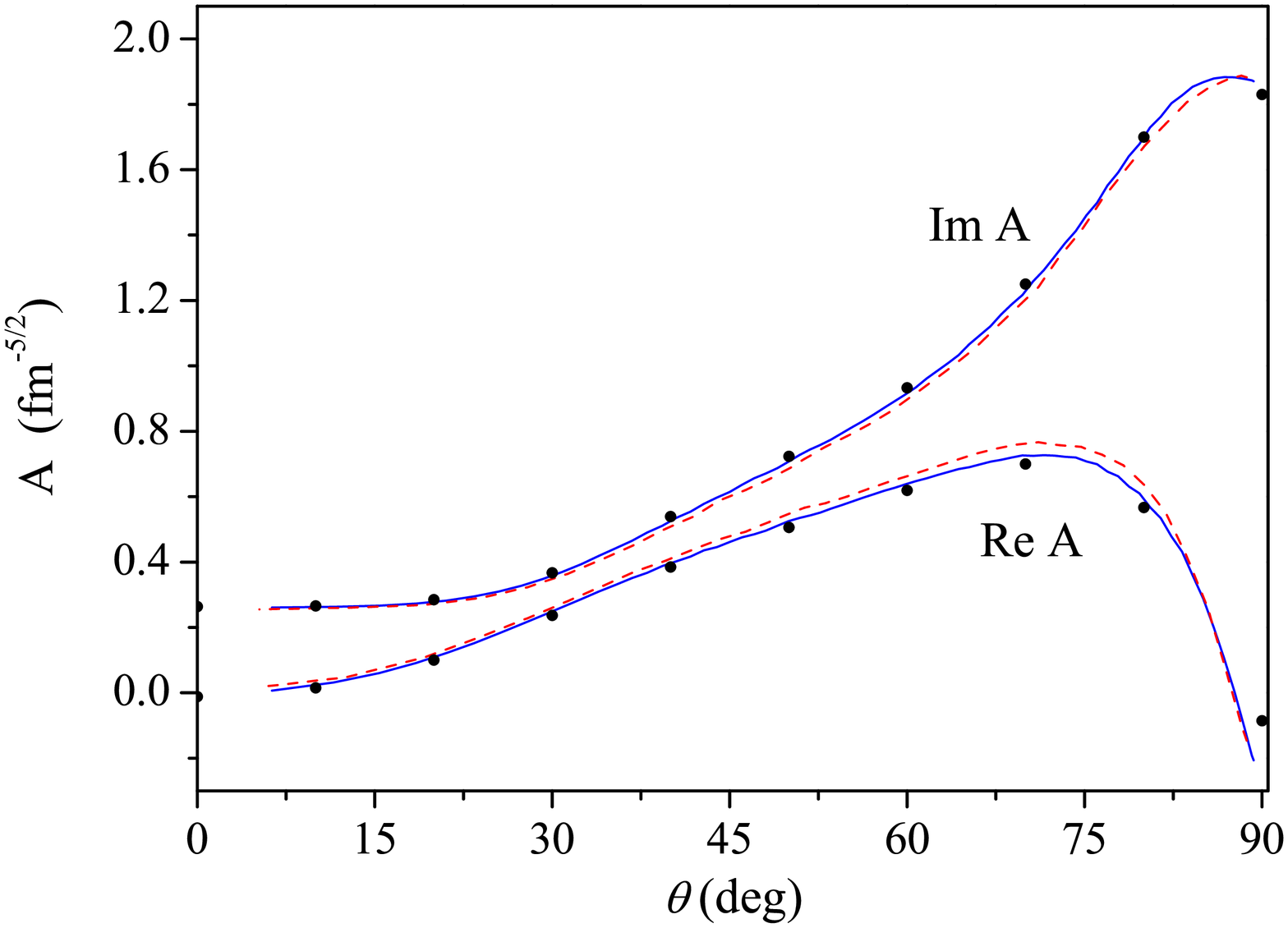,width=\columnwidth}}
 \caption{(Color online) The breakup  ${\cal A}^{\half 1}$ amplitudes in spin-doublet channel for the MT I-III $NN$ potential.
The notations are the same as in  Fig.~\ref{fig4}. \label{fig5}}
 \end{figure}

\begin{figure}[h] \centering
{\epsfig{file=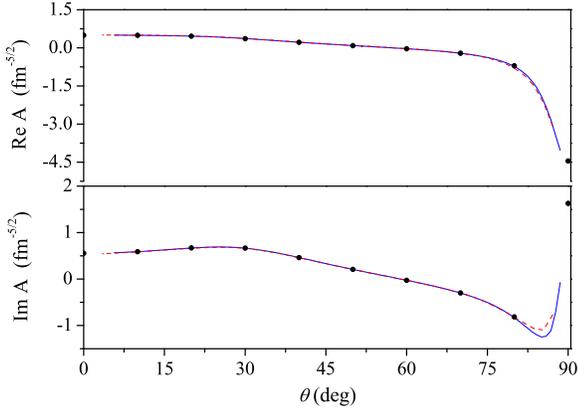,width=\columnwidth}}
 \caption{(Color online) The breakup  amplitudes ${\cal A}^{\half 0}$  in the spin-doublet channel for the  MT I-III $NN$ force.
The notations are the same as in  Fig.~\ref{fig4}. \label{fig6}}
 \end{figure}

The results of such a comparison are presented in
Fig.~\ref{fig4}-\ref{fig6} for our single-component hyperspherical
amplitudes ${\cal A}^{\Sigma s}$. One can observe in the
Figs.~\ref{fig4}-\ref{fig6} a quite satisfactory general agreement
with the results of the benchmark calculations \cite{Friar} except
in the region $\theta\sim$ 90$^\circ$, similarly to the case of the
separable $NN$ interaction. It should be mentioned, that the  from the
   amplitude ${\cal A}$ Eq.~(\ref{ca_mt}) has an additional factor
inversely proportional to the relative momentum $p$ compared to
the amplitude $\mathbb T$ discussed in the previous subsection. So,
the differences to the exact solution of ${\cal A}$  are
more visible at the region corresponding to small values of $p$.

 Similar difficulties  at $\theta\sim$ 90$^\circ$  are also
 observed in other works \cite{Bound_Gl,Vlach,GPF} in which the breakup
calculations have been done in the configuration space. In Ref.~\cite{GPF} it
has been demonstrated that for the correct calculation of the breakup amplitudes
in the area of small relative $NN$ momenta one has to employ the explicit
integral form for the breakup amplitude $T(p,q)$.

In our approach the minor disagreement of the WP breakup
amplitudes and the exact benchmark results at very low relative momenta
can be related to some uncertainties  in the determination of the widths and the
corresponding values of momenta for the WP scattering states of the
pair continuum  very close to the threshold. So the
case of very low relative momenta in the lattice approach deserves 
a separate study which is under way.

\begin{figure}[h] \centering
{\epsfig{file=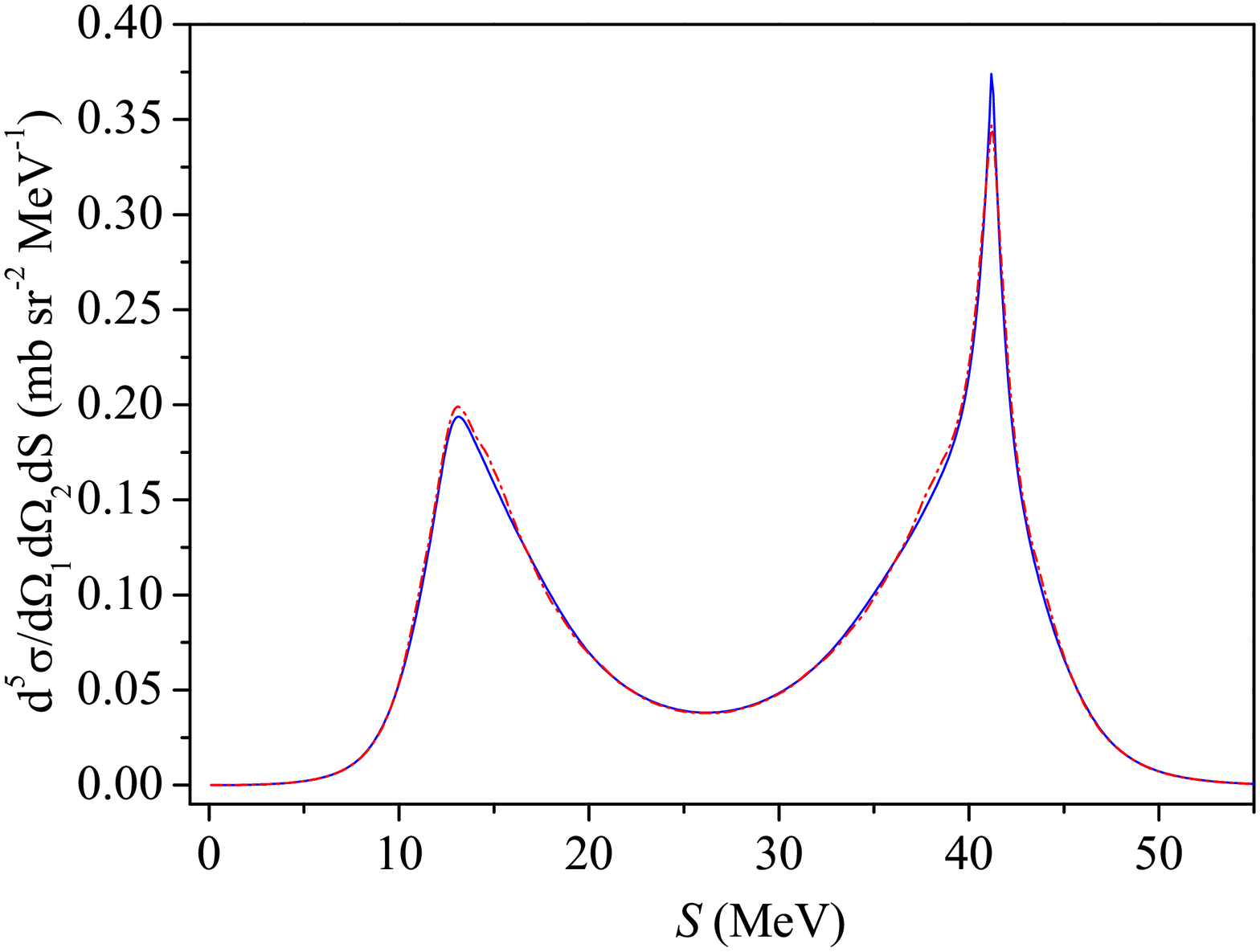,width=\columnwidth}}
 \caption{(Color online) The differential cross section for two-neutron emission at the kinematical configuration:
 $\theta_1=45^\circ$, $\theta_2=50.64^\circ$,
 $\phi_{12}=180^{\circ}$ at the incident neutron energy $E_{\rm lab}=42$~MeV
 found with usage of the WP
technique (dotted curve) and the conventional Faddeev calculations
(solid curve). (''Final state peak''.) \label{fig7}}
 \end{figure}

\begin{figure}[h] \centering
{\epsfig{file=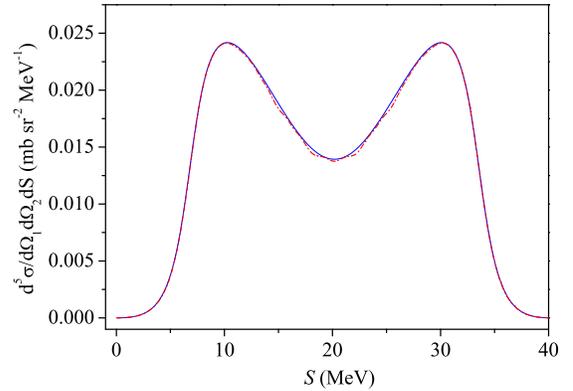,width=\columnwidth}}
 \caption{(Color online) The same as in the Fig.~\ref{fig7} but for the kinematical configuration
 $\theta_1=53.6^\circ$, $\theta_2=53.6^\circ$,
 $\phi_{12}=120^{\circ}$. (''Space star''.) \label{fig8}}
 \end{figure}

Using further the single-component amplitudes we have found the
total (i.e. with inclusion of all three Faddeev components) breakup
amplitudes  as well  as differential cross sections of two-neutron
emission for different kinematical configurations. In
Figs.~\ref{fig7} and \ref{fig8} the differential cross sections
for two-neutron emission with our WP technique are presented
for two configurations and compared to the results of
Ref.~\cite{Friar}. One configuration includes the FSI peak while the
second one is related to the so called ``space-star'' breakup
kinematics. For derivation of the Faddeev cross sections we used an
interpolation of the data in the table presented in
Ref.~\cite{Friar} for the single-component amplitudes.

In summary, the agreement between conventional and lattice
results is generally very good. The curves are mostly hardly distinguishable
in Figs.~\ref{fig7},\ref{fig8}.

\section{Summary}

In the present work we generalized the wave-packet  method developed
by the present authors earlier for  the discretization of the three-body
continuum and used it  for finding the three-body breakup
amplitudes.  As far as the present authors are aware, the  study in
the work is the first where the Faddeev breakup amplitudes are
obtained  completely in the three-body $L_2$-basis. Thus, it would
be appropriate to enumerate some important distinctive features of
our lattice-like approach.

1. Due to projection of the scattering integral equations onto the
wave-packet $L_2$ basis corresponding to the three-body channel
Hamiltonian $H_1$, we get an explicit analytical representation for
the three-body channel resolvent $G_1$ that is used in all further
calculations. For this we employ the version of  the integral
Faddeev equations with the kernel $Pv_1G_1$ instead of the
conventional form  $Pt_1G_0$. This simplifies drastically the whole
calculations scheme as compared to the conventional one, because
first we do not need to know the full off-shell pair $t$-matrix at
many different energies and in addition we get the matrix kernel in a very
convenient finite-dimensional form. As an input information for the
 $NN$-interaction, we use only the results of {\em a single
diagonalization} (for every spin channel) of the $NN$ Hamiltonian
matrix. From such a diagonalization we get immediately the whole set
of pseudo-states (the scattering WP's) and partial phase shifts at
many energies corresponding to these pseudo-states~\cite{KPRF}.

2. For the matrix of the transition operator, one gets an  {\em universal} linear
matrix equation with finite matrix elements. The diagonal (on-shell) elements
of this solution determines the elastic scattering amplitudes while the
non-diagonal (half-shell) elements determines the single-component breakup
amplitudes up to some known phase factor.

3. The structure of the kernel for the matrix equation obtained is
very convenient for numerical realization. Due to the fact that the
kernel is a product of a diagonal matrix, two block  matrices and a
very sparse matrix, it is possible to greatly reduce requirements for
the RAM storage size and noticeably decrease the computation time.

 4. The effect of
the particle permutation operator in the Faddeev kernel  is
represented now with the help of the universal matrix of basis
functions overlapping for different Jacobi coordinate sets. It allows
to avoid very time-consuming numerous re-interpolations of the
current solutions (at iterations) from one set of Jacobi coordinates
to another one at every iteration step \cite{Gloeckle_rep}.

5. Due to an averaging of the integral kernels over the cells in momentum space
the very complicated energy singularities of the kernel above the breakup
threshold (e.g. the moving branching points etc.) are smoothed and one
can solve the few-body scattering equations directly at real energies, i.e.
without any contour deformation to the complex-energy plane. This fact also
facilitates enormously the practical solution of few-body equations above the
breakup threshold.

6. The comparison of the results obtained in our approach with those for the
model for the separable $NN$ potential and with benchmark breakup calculations
(with a semi-realistic local $NN$-potential) has demonstrated that the WP method
allows to get quite accurate three-body breakup amplitudes and cross sections.
Still, the region of very low relative $NN$ momenta requires some additional
study. Some inaccuracy of our results in this area can be related to two
factors: (i) a slow convergence of the WP amplitudes in this region, and  (ii)
some uncertainty of the WP representation of the two-body continuum  in the region
of very low relative momenta. We plan to devote a special study for the solution of
this problem.

In summary one can conclude that the total lattice-like $L_2$
discretization of the three-body continuum allows to find an
accurate  solution for the three-body Faddeev equations for breakup
amplitudes and  simplifies enormously the calculation
together with a noticeable reduction of the computational cost.

{\bf Acknowledgments} The authors appreciate a partial  financial
support from the  RFBR grants
10-02-00096, 10-02-00603, and 12-02-00908.

\appendix
\section{The permutation matrix in the lattice basis}
In our approach we employ a lattice basis, i.e.  a basis
built by free WP's in momentum space. The two-dimensional
(three-body)  wave packet in momentum space are
step-like functions of variables $p$ and $q$:
\begin{equation}
\label{fwp} \langle p,q |\mathfrak{p}_i\mathfrak{q}_j\rangle
\equiv\langle p,q |\MD_{ij}\rangle =
\frac{1}{\sqrt{d_{i}\bar{d}_{j}}}{\vartheta(p\in\MD_i)}
{\vartheta(q\in\BMD_j)},
\end{equation}
which are nonzero only at the intervals $\MD_i=[p_{i-1},p_i]$
 and $\BMD_j=[q_{j-1},q_j]$
($d_{i}$ and $\bar{d}_{j}$ are the widths of
 corresponding intervals).
Such wave packets are normalized to
unity with the weight $dpdq$ and form an orthonormal basis (it is assumed
that the intervals are not overlapping).

The matrix element of the permutation operator  $P$ between
plane waves has a simple form for the  $s$-wave:
\begin{eqnarray}
\label{P0} \langle p',q',s'\Sigma|P|p,q,s\Sigma\rangle
\equiv\Lambda^\Sigma_{s's} P^0(p',q',p,q)=
\nonumber \\
\Lambda^\Sigma_{s's}\  {4}\delta ({p'}^2+\trelf{q'}^2- (p^2+\trelf
q^2))\, \vartheta(1-|x|),
\end{eqnarray}
where $\Lambda^\Sigma_{ss'}$ is a spin-channel coupling matrix.
The $\delta$-function guaranties  energy conservation and $x$ is the
cosine of the angle between vectors $\bf q$ and $\bf q'$ which (with
taking into account the $\delta$-function) can be expressed as a
function of three momenta, e.g. $p$, $q$, $q'$:
\begin{equation}
x=\frac{p^2-{q'}^2-q^2/4}{qq'}.
\end{equation}
The condition $|x|<1$ in (\ref{P0}) restricts the allowed values of momenta
to a region, where the overlap is nonzero.

To find the matrix elements of the permutation operator $P$ over the
free WPs (\ref{fwp}),
 one has to integrate the  function  $ P^{0}( p',q',p,q)$
over rectangular cells $\MD_{ij}=\MD_i\otimes\BMD_j$,
$\MD'_{i'j'}=\MD'_{i'}\otimes\BMD'_{j'}$ (where the upper prime at
the interval symbol denote that it  refers  to a different
set of Jacobi coordinates):
\begin{eqnarray}
\label{p_pack}
 \langle \MD'_{i'j'} |P|\MD_{ij}\rangle = \frac{1}
 {\sqrt{d_{i}\bar{d}_{j}d_{{i'}}\bar{d}_{{j'}}}} \times \\
 \int_{\MD'_{i'j'}}\int_{\MD_{ij}}P^{0}( p',q',p,q)\,dp\,dq\,dp'\,dq'.
 \nonumber
\end{eqnarray}
Actually, the integral in Eq.~(\ref{p_pack})  is reduced to an
 area  of two overlapping  rectangular areas $\MD_{ij} $
and $\MD'_{i'j'}$.

Hyperspherical (polar in the s-wave case) coordinates are most convenient  to
calculate such overlaps. Let us introduce the reduced (rescaled) momentum
variable $\tilde{q}$:
\begin{equation}
\tilde{q}=\sqrt{(3/4)}q,
\end{equation}
then, the energy conservation takes the ``homogeneous'' form
$p^2+\tilde{q}^2={p'}^2+{\tilde{q}}^{'2}$.
 The hyperspherical
coordinates $Q,\alpha$ are introduced as usually:
\begin{equation}
\tilde{q}=Q\sin\alpha,\quad p=Q\cos\alpha,\quad Q^2=p^2+\tilde{q}^2.
\end{equation}
In these hyperspherical coordinates the integral in Eq.~(\ref{p_pack})  takes the following
form:
\begin{align}
\int\delta ({p'}^2+\trelf{q'}^2-(p^2+\trelf q^2))
\vartheta(1-|x|)\,dp\,dq\,dp'\,dq'\nonumber \\
=(4/3)\int\delta (Q^2-{Q'}^2)\vartheta(1-|x|)\,QdQd\alpha Q'dQ' d\alpha'\nonumber \\
= 1/3\Pi(\MD_{ij},\MD'_{i'j'})\label{p00},
\end{align}
where we define the overlapping square:
\begin{equation}
\Pi(\MD_{ij},\MD'_{i'j'})\equiv\int\vartheta(1-|x|)\,d(Q^2)d\alpha
d\alpha'.
\end{equation}
 Thus we get that the permutation matrix element is directly interrelated to this square:
\begin{equation}
 \langle \MD'_{i'j'} |P|\MD_{ij}\rangle =
 \frac{4}{3}\frac{\Pi(\MD_{ij},\MD'_{i'j'})}
 {\sqrt{d_{i}\bar{d}_{j}d_{i'}\bar{d}_{j'}}}.
\label{pDD}
\end{equation}
The condition $|x|<1$ can be expressed through the hyperangular variables
$\alpha, \alpha'$ as follows:
\begin{equation}
\label{ineq}
|\frac{\pi}{3}-\alpha|<\alpha'<\frac{\pi}{2}-|\alpha-\frac{\pi}{6}|
\end{equation}
So, the
 overlap region
 $S(\alpha, \alpha')$  determined by the condition $|x|<1$ is
a rectangle in the plane $(\alpha , \alpha')$ restricted by four straight lines (see Fig.~\ref{fig9}):

\begin{figure}[h]
\begin{center}
\epsfig{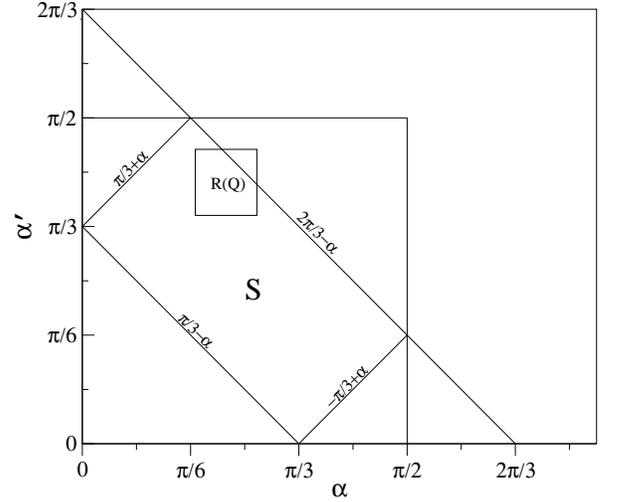}
\end{center}
\caption {The integration region in the plane ($\alpha,\alpha'$)
which is the intersection of the  area of allowable values of
$\alpha,\alpha'$ --- the large rectangle  $S$ determined by the
inequalities (\ref{ineq}) --- and the  rectangle $R(Q)$ which
boundaries depend on the value of $Q$.} \label{fig9}
\end{figure}

Therefore, the integral in Eq.~(\ref{p00}) can be evaluated as
the external (numerical) integral over
$Q^2$ in the range between $Q^2_{\rm min}$ and $Q^2_{\rm max}$ from
the area of intersection of the rectangle $S$ and
the rectangle $R(\alpha_{\rm min}, \alpha_{\rm max},\alpha'_{\rm
min},\alpha'_{\rm max})$ whose vertices depend on $Q$  (see Fig.~\ref{fig9}):
\begin{equation}
\Pi=\int_{Q^2_{\rm min}}^{Q^2_{\rm max}}d(Q^2)\int\int_{S\cap
R(Q)}d\alpha d\alpha'.
\end{equation}
The integration limits  over $Q^2$ are equal:
\begin{align}
Q^2_{\rm min}=\max(p_{i-1}^2+\tilde{q}_{i-1}^2,
{p'}_{j-1}^2+(\tilde{q}'_{j-1})^2),
\label{Q1}\\
Q^2_{\rm max}=\min(p_{i}^2+\tilde{q}_{i}^2,
{p'}_{j}^2+(\tilde{q}'_{j})^2).
\label{Q2}
\end{align}
If $Q^2_{\rm max}<Q^2_{\rm min}$ then the cells do
not overlap and the matrix element is equal to 0.

\begin{figure}[h]
\begin{center}
\epsfig{file=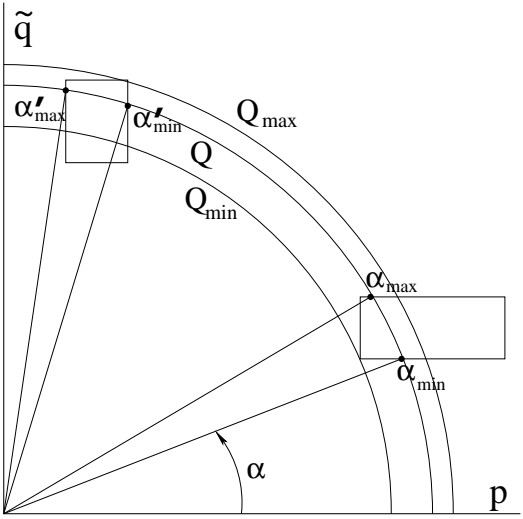,width=0.8\columnwidth}
\end{center}
\caption {On the definition of the integration limits   (\ref{Q1}) -
(\ref{Q4}) in the variables $Q,\alpha, \alpha'$. The cells
$\MD_{ij}$ and $\MD'_{ij}$ in the plane ($p, \tilde{q}$) and their
polar coordinates $(\alpha_{\rm min},\alpha_{\rm max})$ and
$(\alpha'_{\rm min},\alpha'_{\rm max})$ are shown.} \label{fig2n}
\end{figure}

The coordinates of vertices of the rectangle
$R(Q)$ are computed directly (see the Fig.~\ref{fig2n} for further
explanations):
\begin{align}
\alpha_{\rm
min}(Q)=\max(\arcsin\frac{\tilde{q}_{i-1}}{Q},\arccos\frac{p_i}{Q});\nonumber\\
\alpha_{\rm max}(Q)=\min(\arcsin\frac{\tilde{q}_{i}}{Q},\arccos\frac{p_{i-1}}{Q});\label{Q3}\\
\alpha'_{\rm
min}(Q)=\max(\arcsin\frac{\tilde{q}'_{j-1}}{Q},\arccos\frac{p'_j}{Q});\nonumber\\
\alpha'_{\rm
max}(Q)=\min(\arcsin\frac{\tilde{q}'_{j}}{Q},\arccos\frac{{p'}_{j-1}}{Q}).\label{Q4}
\end{align}
Here, if $p_i>Q$ then  $\arccos\frac{p_i}{Q}$ should be replaced by
0, and if $\tilde{q}_i>Q$ then $\arcsin\frac{\tilde{q}_i}{Q}$ should
be replaced by $\pi/2$; the same rule should be applied also to
primed values.

The area of intersection $S\cap R(Q)$ in the plane
$\alpha, \alpha'$ is evaluated analytically by the formulas of
elementary geometry.

\section{Wave-packet solution for the  three-body scattering problem with a separable potential}
We consider here a system of three nucleons
 with equal masses $m$ interacting by a separable $NN$ force
 in  s-wave spin-singlet ($s=0$) and spin-triplet
($s=1$) states correspondingly:
\begin{equation}
\label{seppot} v_s=\lambda_s|\varphi_s\rangle\langle\varphi_s|,\quad
s=0,1.
\end{equation}
The two-body $t$-matrices are also separable:
\begin{equation}
t_s(E)=|\varphi_s\rangle\tau_s(E)\langle\varphi_s|, \label{sepp}
\end{equation}
where $\tau$ are known functions
\begin{equation}
\tau^{-1}_{s}(E)=
\lambda_{s}^{-1}-\langle\varphi_{s}|g_0(E)|\varphi_{s}\rangle,
\end{equation}
and $g_0(E)$ is the free two-particle resolvent.  Function $\tau_1$
for triplet state has the pole at the deuteron binding energy
$\vep_0^*$. The corresponding bound state wavefunction
$|z_0\rangle$ is defined as follows:
\begin{equation}
|z_0\rangle=\sqrt{R(\vep_0^*)}g_0(\vep_0^*)|\varphi_1\rangle,
\label{psisep} \end{equation} where $R(\vep_0^*)$ is the residue of
$\tau_t(E)$ at the pole.

We  use here the two-parameter  Yamaguchi potentials with
the form factors:
\begin{equation}
\varphi_{s}({p})=(p^2+\beta_{s}^2)^{-1}.
\end{equation}
In this case  $\tau_s(E)$ and $R(\vep_0^*)$ take the form:
\begin{eqnarray}
\tau^{-1}_s=\left(\lambda_{s}^{-1}+\frac{\pi
m}{\beta_s}\frac{1}{(\beta_s -i\sqrt{mE})^2}\right),
\nonumber\\
 R(\vep_0^*)=\frac{\beta_1(b^2+p_b^2)^3p_b}{\pi m^2}, \quad p_b=\sqrt{-m\vep_0^*}.
\end{eqnarray}
The potential parameters $\beta_s$ and $\lambda_s$ are taken from
Ref.~\cite{Cahill}.

The Faddeev equation  for the elastic transition operator $U$ in this
case reduces to the system of
one-dimensional integral equations  of the Lippmann--Schwinger
type~\cite{schmid} for the elastic scattering amplitudes
corresponding to the total spin $\Sigma=\half, \tralf$ and orbital
momentum $L$ (in the case of the $s$-wave pair interactions the $\Sigma$ and $L$
are conserved separately):
 \begin{eqnarray}
F^{\Sigma L}_{ss'}({q,q'};E)=Z^{\Sigma
L}_{ss'}({q,q'};E)+\sum_{s^{\prime\prime}} \int
(q^{\prime\prime})^2{ d}q^{\prime\prime}\times
\nonumber\\
 Z^{\Sigma L}_{ss^{\prime\prime}}({ q',q^{\prime\prime}};E)
\tau_{s_1}\left(E-\frac{3({q^{\prime\prime}})^2}{4m}\right)
F^{\Sigma L}_{s^{\prime\prime}s'}({q^{\prime\prime},q'};E),\quad
\label{vecfe}
\end{eqnarray}
where
 \begin{equation}
 F_{ss'}^{\Sigma L}({q,q'};E)\equiv \langle{q},\varphi_{s},\Sigma L|
g_0(E)U(E)g_0(E)|{q'},\varphi_{s'},\Sigma L\rangle  .
\end{equation}

The kernels $Z^{\Sigma L}_{ss'}({ q,q'};E)$ in
Eq.~(\ref{vecfe})  are defined as follows:
 \begin{eqnarray}
Z^{\Sigma L}_{ss'}({ q,q'};E)=\nonumber \\
\Lambda^\Sigma_{ss'}2\pi\int_{-1}^{1}
dx\,P_L(x)\,\frac{\varphi_s({\bf q'}+{\bf q}/2)\varphi_{s'}({-\bf
q}-{\bf q'}/2)} {E-q^2/m-{q'}^2/m-{\bf qq'}/m} \label{Zvec}
\end{eqnarray}
where $x=\cos(\widehat{\bf{qq'}})$, $P_L$ are the Legendre
polynomials, and $\Lambda^\Sigma_{ss'}$ is a spin-channel coupling
matrix. In case of quartet scattering one has the single equation
with $s=1$ and $\Lambda^{\tralf}_{11}=-1$, while there are two
coupled equations with $s=0,1$ in the case of doublet scattering and
$\Lambda^{\half}_{00}=\Lambda^{\half}_{11}=\half$,
$\Lambda^{\half}_{01}=\Lambda^{\half}_{10}=-\tralf$

After solving  Eq.~(\ref{vecfe}), the partial wave elastic on-shell
amplitude can be defined from the relation
\begin{equation}
A^{\Sigma L}_{\rm el} (E)=\frac{2m}{3}q_0R(\vep_0^*)F^{\Sigma
L}_{11}({ q_0,q_0};E), \label{ampl}
\end{equation}
where $q_0=\sqrt{\fotrelf  m(E-\vep_0^*)}$.

 Now we turn to the determination of the breakup amplitude.
Substituting the explicit formulas for  the $t$-matrix in  (\ref{sepp}) and
the deuteron wave function (\ref{psisep}) into  (\ref{abr}), one can
express the partial ``Faddeev'' breakup amplitudes via the elastic
amplitudes $F^{\Sigma L}_{ss'}$:
\begin{equation}
T^{\Sigma
L}_s(p,q)=\sqrt{R(\vep_0^*)}\varphi_s({p})\tau_s(E-3q^2/4m)F^{\Sigma
L}_{s1}(q,q_0;E).
\end{equation}

Let's now proceed with the  lattice version for
elastic amplitude. We introduce the free WP basis (\ref{iq}) and
project  Eq.~(\ref{vecfe}) to this basis. Finally, we find the
 matrix equation:
\begin{equation}
{\mathbb F}={\mathbb Z}+{\mathbb Z}{\mathbb \tau}{\mathbb F},
\end{equation}
where the letters with double lines denote matrices of corresponding operators
in the  WP subspace for the given values of total spin $\Sigma$ and
orbital momentum $L$ (below we shall omit $\Sigma$ and $L$ for
brevity). More definitely,
\begin{eqnarray}
Z^{ss'}_{jj'}=\frac{1}{\sqrt{\bar{d}_j\bar{d}_{j'}}}\int_{\BMD_{j}\BMD_{j'}}
dqdq'Z^{\Sigma L}_{ss'}(q,q';E) \nonumber\\
F^{ss'}_{ii'}=\frac{1}{\sqrt{\bar{d}_j\bar{d}_{j'}}}\int_{\BMD_{j}\BMD_{j'}}
dq dq'F^{\Sigma L}_{ss'}(q,q';E)\nonumber \\
\tau^s_{j}=\frac{1}{\sqrt{\bar{d}_j}}\int_{\BMD_{j}} dq q^2
\tau_s(E-3q^2/4m) (q,q';E)\nonumber.
\end{eqnarray}
The elastic on-shell amplitude in WP representation is defined
from the diagonal ``on-shell'' matrix elements of the lattice
transition matrix $\mathbb X$:
\begin{equation}
A_{\rm el}(E)\approx
\frac{2m}{3}q_0\frac{R_b(\vep_0^*)F^{11}_{j_0j_0}}{\bar{d}_j},\quad
E-\vep_0^*\in \BMD_{j_0}
\end{equation}
 Similarly, the packet approximation for the breakup amplitude is determined by
off-diagonal matrix elements of $\mathbb F$:
\begin{equation}
T_s({ p},{
q})\approx\frac{\sqrt{R(\vep_0^*)}\varphi_s({p})\tau_s(q_j^*)F^{s1}_{jj_0}}{\sqrt{\bar{d}_j\bar{d}_{j_0}}},\quad
q\in{\BMD_j},
\end{equation}
where $\tau_s(q_j^*)=\tau_s(E-3{q_j^*}^2/4m)$ and $q_j^*$ is the
midpoint of bin $\BMD_j$.

The WP representation for the breakup amplitude ${\cal A}(\theta)$ which
determines the asymptotics of the breakup wave function in
hyperspherical coordinates has the form:
\begin{equation}
{\cal A}_s(\theta)=\frac{4\pi}{3\sqrt{3}}
\frac{m}{\hbar^2}e^{i\pi/4}K^4q_0\frac{\sqrt{R(\vep_0^*)}\varphi_s({
p})\tau_s(q_j^*)F^{s1}_{jj_0}}{\sqrt{\bar{d}_j\bar{d}_{j_0}}}.
\end{equation}

\section{Features of  the numerical procedure}

Here we will discuss some details of the numerical procedure
for solving the matrix equation~(\ref{m_pvg}). The main difficulty
is its large dimension. Quite satisfactory results can be obtained
with a  basis size $M\sim N\sim 200$. It is means that in the simplest
one-channel (quartet) case one gets a kernel matrix with dimension $M\times
N\sim 40000\times 40000$ which occupies $\sim 6.4$~GB (at single precision) of
RAM  or external memory of the  computer. In the two-channel (doublet) case the required
amount of memory increases by a factor 4.

However, the matrix of the kernel $\mathbb K$ for equation~(\ref{m_pvg}) can be
written as the
product of four matrices which have the
specific structure:

\begin{equation}
\label{kernel}
\mathbb K=\mathbb P {\mathbb V}_1  \mathbb G_1
\equiv \mathbb O \mathbb P^0 \tilde{\mathbb V}_1  \mathbb G_1,
\end{equation}
where
\[\tilde{\mathbb V}_1 =  \mathbb O^{\rm T}\mathbb V_1. \]
The matrix of the channel resolvent $\mathbb G_1$  is diagonal and its elements are
defined by simple explicit formulas. The matrix of the
potential  $\mathbb V_1$ has a block-type structure (\ref{bv1}): in fact, it is the direct
product of the $(M\times M)$-matrix of the  two-body interaction and the unit
$(N\times N)$-matrix. The rotation matrix $\mathbb O$ has a  similar form and
the actual dimension $(M\times M)$. The free permutation matrix $\mathbb P^0$
is very sparse due to the energy conservation
 condition. As a rule, only about ~$1\%$ of
its elements are distinguished from zero, and the sparsity increases
when the basis dimension increases.

So, if to summarize all these details one finds that the kernel in
the matrix equation  (\ref{m_pvg}) is the product of four matrices:
a diagonal one, a very sparse one and two block matrices with
actual dimension $(M\times M)$. If,  instead of storing the entire
matrix $\mathbb K$, we store  its factors only (for
the sparse matrix $\mathbb P^0$ we store the nonzero elements only),
then we can save a huge portion of physical memory: for the above
example we shall need only $\sim 128$~MB instead of initial 6.4~GB.
Such an enormous reduction
 of the required memory allows us to perform
calculations without using an external memory, which, in its turn, reduces the calculation time
by approximately one order.

This  possibility  to avoid storing a very large amount of data is
related to the specific procedure used by us for the solution of the
equation (\ref{m_pvg}). As a matter of fact, to find the elastic and
breakup amplitudes one needs not all but only on-shell matrix
elements of the transition operator. So, each of these elements can
be found without complete solving the matrix equation (\ref{m_pvg})
but by means of a simple iteration procedure  with subsequent
summing the iterations via the Pade-approximant technique. If we do
not store the entire matrix of kernel $\mathbb K$, to perform
each iteration we need only three additional matrix-vector
multiplications with the matrices of a  special form.

All these features of the procedure  lead to an  extremely economic
calculation scheme which can be realized with a usual moderate PC.

\end{document}